\documentclass[aps,twocolumn,superscriptaddress,a4paper]{revtex4}
\usepackage{amsfonts,amssymb,amsmath}
\usepackage[utf8]{inputenc}
\usepackage[pdftex,breaklinks=true,colorlinks=true]{hyperref}
\usepackage[pdftex]{graphicx}
\usepackage{epstopdf}
\usepackage{leftidx}
\graphicspath{{Pictures/}}

\newcommand{\ket}[1]{\left|#1\right>}
\newcommand{\bra}[1]{\left<#1\right|}

\usepackage{xcolor}
\usepackage{float}

\begin{document}

\title{Out-of-equilibrium transport in  the  interacting resonant level model: the surprising relevance of the  boundary sine-Gordon model}

\author{Kemal Bidzhiev}
\affiliation{Institut de Physique Th\'eorique, Universit\'e Paris Saclay, CEA, CNRS, F-91191 Gif-sur-Yvette, France}
\author{Gr\'egoire Misguich}
\affiliation{Institut de Physique Th\'eorique, Universit\'e Paris Saclay, CEA, CNRS, F-91191 Gif-sur-Yvette, France}
\affiliation{Laboratoire de Physique Théorique et Modélisation, CNRS UMR 8089, Université de Cergy-Pontoise, F-95302 Cergy-Pontoise, France}
\author{Hubert Saleur}
\affiliation{Institut de Physique Th\'eorique, Universit\'e Paris Saclay, CEA, CNRS, F-91191 Gif-sur-Yvette, France}
\affiliation{Department of Physics and Astronomy, University of Southern California, Los Angeles, CA 90089-0484, USA}

\date{\today}

\begin{abstract}
Using time-dependent density matrix renormalization group calculations we study the transport properties ($I-V$ curves and shot noise) of  the interacting resonant level model (IRLM)
in a large range of the interaction parameter $U$, in the scaling limit. We find that these properties can be described remarkably well by those of the Boundary sine-Gordon model (BSG), which  are known analytically (Fendley, Ludwig and Saleur, 1995). We argue that the  two models are nevertheless  in different universality classes out of equilibrium: this requires a  delicate discussion of their infra-red (IR) properties ({\it i.e.} at low voltage), where we prove in particular 
 that the effective tunneling charge is $e$ in the infra-red regime of the IRLM (except at the self-dual point where it jumps to $2e$), while it is known to be a continuously varying function of $U$ in the BSG. This behavior is confirmed by careful analysis of the numerical data in the IR. 
The remarkable agreement of the transport properties, especially in the crossover region, remains however unexplained. 

\end{abstract}

\maketitle


\section{Introduction}

The field of out-of-equilibrium quantum many body systems has developed rapidly in the last decade.
From studies about the equilibration in isolated systems~\cite{polkovnikov_colloquium_2011,eisert_quantum_2015},  to questions about transport properties, entanglement entropy, dissipation or the effect of disorder, the issues raised are very diverse, and often fundamental. However, despite the large amount of work  - both analytical and numerical - devoted  to this field, much fewer exact results are available than in the equilibrium case. In particular, the question of how and when a model that is integrable in equilibrium can also be solved out-of-equilibrium remains open.

A notable exception to this state of affairs concerns transport through a weak link in a Luttinger liquid \cite{Kane-Fisher}. This model  can be mapped in the low-energy limit onto the boundary sine-Gordon model (BSG), and, thanks to a variety of techniques \cite{GoshalZamolodchikov, FLS, BLZ}, exact expressions can be obtained for many  stationary transport properties, and even some non-stationary ones \cite{carr_full_2015}. Via exact mappings, the solution of the BSG model also provides results for e.g. transport through point contacts between edges states in the fractional quantum Hall effect, or for the Josephson current in small dissipative superconducting junctions \footnote{All these results hold in the scaling limit, and in this limit only.}

One of the very pleasant  features of the BSG model is that it involves a freely adjustable parameter, which can be interpreted as the anomalous dimension of the tunneling operator at the UV or IR limit of the renormalization group flow, and characterizes, among others, the anomalous power laws in the $I-V$ characteristics. Of course, this parameter takes a fixed value in a given realization - e.g. for a given value of the Luttinger liquid interaction, or the quantum Hall filling fraction.

The BSG model is solvable both in and out-of-equilibrium (at least as far as charge transport is concerned): it is not clear whether there are many other models sharing this nice feature. An obvious candidate where this question can be investigated is the interacting resonant level model (IRLM)~\cite{wiegmann_resonant-level_1978}, whose solution in equilibrium is well known \cite{filyov_method_1980}. For technical reasons involving the nature of the quasiparticles scattering in the Bethe-ansatz, the tricks used in \cite{FLS,BLZ} to solve the BSG model out-of-equilibrium do not work in this case. In ~\cite{boulat_twofold_2008}, a solution out-of-equilibrium was found at a special (self-dual) point $U=U_{\rm sd}$, where a hidden $SU(2)$ symmetry is present, that makes a mapping onto the BSG model  possible. It was   speculated in ~\cite{boulat_twofold_2008} that the IRLM out-of-equilibrium might well be  solvable only at that point (and of course at the point where the interaction vanishes, and the model 
reduces to the usual resonant level model, RLM). 
Meanwhile, in \cite{andrei}, a new ``open Bethe-ansatz'' was proposed, that should in principle allow for a new solution of the model in equilibrium, and also make possible the calculation of, for instance,  the steady current {\sl for all values of the interaction}. Unfortunately, the method proposed in  \cite{andrei}, while not devoid of ambiguities, has also proven impossible to exploit in the scaling regime, where meaningful comparison with numerics could be carried out \footnote{The results in \cite{andrei} are obtained with a cut-off that is not the ordinary lattice cut-off, and thus ``bare data''  cannot be compared with  simulations on an ordinary lattice model.}. The question thus remains open: can the I-V characteristics (for instance) of the IRLM be obtained using integrability techniques for all values of the interaction $U$?

A closer look at the solution of the BSG model shows that many of its non-equilibrium properties could have been guessed (at least at $T=0$) by using general arguments of analyticity and duality \cite{FendleySaleurDuality, FendleySaleurHyperelliptic,FendleySaleurDiffEq}. It is tempting to think that, if the IRLM is indeed solvable out-of-equilibrium, educated guesses can be made for some of its properties, by exploiting the underlying {\sl equilibrium} RG flow, which is of Kondo type. In order to do so of course, it is necessary to have accurate numerical results, for all values of $U$ (or at least a significant range),  to compare with. It is with this goal in mind that we have carried out - as reported in the first part of  this paper - an extensive numerical  (Sec.~\ref{sec:def}) study, of the current (Sec.~\ref{sec:current}) and shot noise (Sec.~\ref{sec:charge_fluct}) in the IRLM at $T=0$. What we found is surprising:  these properties are almost identical with those of the BSG model, once the 
adjustable parameter in the latter is properly chosen to ensure the same  behavior at leading order in the UV. This holds to a remarkable (and largely unexpected)  accuracy, raising the question of whether the solution of the BSG model could, in fact,  also be the solution of the IRLM, at least at $T=0$. We discuss this point in the second half of this paper (Sec.~\ref{sec:IRLMvsBSG}), and answer it in the negative by focusing, both analytically and numerically,  on subtle aspects of the $I-V$ characteristics in the IR limit.
We show in particular, using bosonization {\sl and integrability in equilibrium}, that the quasiparticles which tunnel across the dot have an effective charge $e$ in the IR limit of the IRLM (except at the self-dual point, where it is $2e$), which departs from the continuously varying charge in the BSG model. This is confirmed
by some numerical calculations of the backscattered current and shot noise at low bias in the IRLM at intermediate value of the interaction.
Why the results for the IRLM and the BSG model are so close remains however an open question, which we discuss some more in the conclusion. 

\section{Definition of the model}
\label{sec:def}

\subsection{Hamiltonian}\label{ssec:model}
The IRLM can be defined in terms of spinless fermions on a one-dimensional lattice:
\begin{eqnarray}
 &H_{\rm IRLM}&=H_L+H_R +H_d \label{eq:H_IRLM} \\
 &H_L&=-J\sum_{r=-N/2}^{-2} \left(c^\dagger_r c_{r+1} +{\rm H.c}\right) \\
 &H_R&=-J\sum_{r=1}^{N/2-1} \left(c^\dagger_r c_{r+1} +{\rm H.c}\right) \\
 &H_d&=-J' \sum_{r=\pm 1} \left( c^\dagger_{r} c_{0} +{\rm H.c}\right) +\nonumber\\
 &&+ U\sum_{r=\pm 1} \left(c^\dagger_r c_r-\frac{1}{2}\right)\left(c^\dagger_0 c_0-\frac{1}{2}\right).
 \end{eqnarray}
 \begin{figure}[h!]
\includegraphics[width=0.5\textwidth, angle=0]{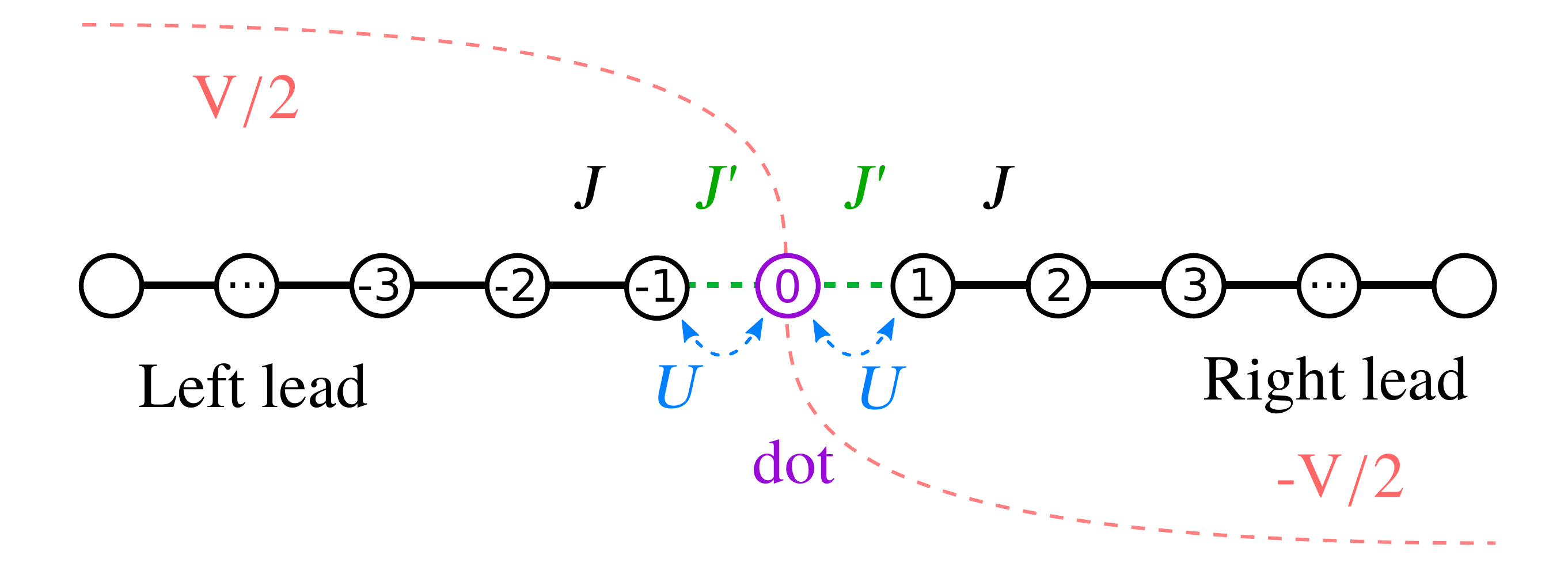}
\caption{Schematics of the IRLM. The system is prepared at $t=0$
in the ground state of the model with a chemical potential $V/2$ in the left lead, and $-V/2$ in the right lead.
For $t>0$ the system then evolves with the bias switched to zero.}
\label{fig:system}
\end{figure}	
$H_L$ and $H_R$ describe fermions hopping (or kinetic energy) in the left and the right ``leads'',
and $H_d$ encodes the tunneling from the leads to the dot (level at $r=0$) and the density-density interaction (strength $U$) between the dot and the leads.
In the following we set the unit of energy to be the hopping amplitude $J=1$ (bandwidth in the leads equal to $W=4J=4$).

In what follows we are interested in the so-called {\it scaling regime} where the bandwidth is much larger that the other energies in the problem. In this regime many properties can be calculated using the continuum counterpart, modulo the replacement of $U$ (lattice model) by $U_c$ (in the continuum, see Sec.~\ref{ssec:large_bias}).
Strictly speaking the scaling regime is attained in the lattice model when $0<J'\ll J$ and $V\ll W$.
On the other hand, the numerics are necessarily performed with a finite $J'$ and finite $V$, and too small values of these parameters would lead to large transient times,
exceeding the times accessible to the simulations. One should therefore work in a range of bias and hopping such that, at the same time, 
corrections to scaling are negligible and a quasi-steady values are reached in a time accessible to the numerics. 
As observed in previous studies on this problem (see for instance \cite{boulat_twofold_2008,bidzhiev_out-equilibrium_2017}), it is fortunately possible
to satisfy these constrains.
Checking that the numerical data indeed correspond to the scaling regime is a important part of the work, and some discussion of this point is presented in Appendix \ref{sec:numerics}.
We also come back to this when discussing the numerical results.

As previously done in several works~\cite{schmitteckert_nonequilibrium_2004,schneider_conductance_2006,boulat_twofold_2008,branschadel_conductance_2010,carr_full_2011,bidzhiev_out-equilibrium_2017}, the initial state is prepared with an inhomogeneous particle density, using a bias voltage $+V/2$ in the left, and $-V/2$ in the right one.  Note that instead of using a step function, we use a smooth function interpolating between $+V/2$ and $-V/2$:
\begin{equation}
 H_{\rm bias} = V/2\sum_{r=-N/2}^{N/2} \tanh(r/w) c^\dagger_r c_{r}
 \label{eq:tanh}
\end{equation}
where $2w$ is a smoothing width. We typically take $w=10$ lattice spacings.

\subsection{Quench protocols}
\label{ssec:protocols}

In this work we use and compare two different quench protocols, dubbed (A) and (B). In both cases the bias $V$ is switched to zero for $t>0$, and the wave function evolves according to $\ket{\psi(t)} = \exp(-i H_{\rm IRLM}t)\ket{\psi}$
with some interaction parameter $U$ and tunneling amplitude $J'$ between the leads and the dot.
The two protocols differ by their initial states.
In the protocol (A) the initial state is prepared as the ground state of $H_{\rm bias}+H_{\rm IRLM}(U_0,J'_0)$ with $U_0=0$ (free fermions) and with an homogeneous hopping amplitude $J'_0=J=1$ in the whole chain.

On the other hand, in the protocol (B)  the initial state is the ground state of $H_{\rm bias}+H_{\rm IRLM}(U,J')$. In other words, $U$ and $J'$ are not changed at the moment of the quench. For a simple energy reason (see Appendix~\ref{sec:compare_protocols}), this initial state should be preferred at large $U$. It also produces a lower amount of entanglement entropy and thus allows for longer simulations compared to (A).

\subsection{Current}

The (particle) current flowing through the dot is defined as
\begin{equation}
I(t)=\frac{1}{2} \left[I(-1,t)+I(0,t)\right]
\end{equation}
where
\begin{equation}
I(r,t)=2 J_{r,r+1} \  {\rm Im} \bra{\psi(t)}c^\dagger_r c_{r+1} \ket{\psi(t)} 
\end{equation}
is the expectation value of the current operator associated with the bond $r,r+1$. For the two bonds which connect the dot to the left and right leads
the hopping amplitude is $J_{r,r+1}=J'$ (and $J_{r,r+1}=J$ in the leads).
 This current is expected to reach a steady
value when $t$ is large (but keeping $t v_F$ smaller than the system size, where $v_F=2J=2$ is the Fermi velocity in the leads). This steady value  is extracted from the numerical data by fitting
$I(t)$ to a constant plus damped oscillations (more details in Appendix~\ref{sec:compare_protocols}).

\subsection{Charge fluctuations and shot noise}
\begin{figure}[h!]
  \includegraphics[height=0.5\textwidth, angle =270]{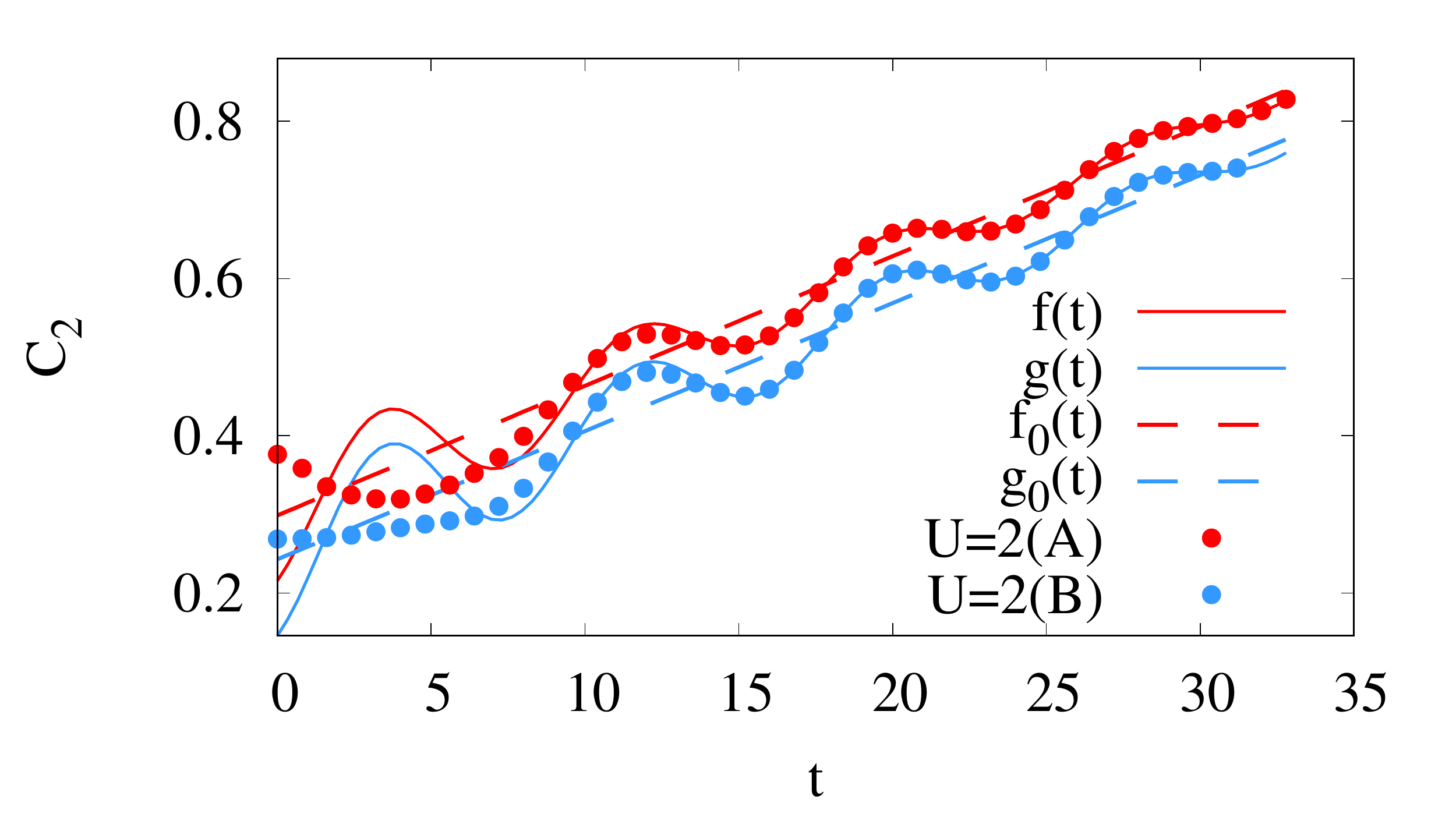}
\caption{Time evolution of the second charge cumulant $C_2$. Red circles represent $C_2$ for the IRLM protocol A, blue circles represent the protocol B with parameters of the model $U=2$ (self-dual point), $J'=0.2$ and $V=1.6$.
The data can be interpreted as a linear growth superposed with (slowly damped) oscillations. The fitting functions are $f(t)=0.3+0.0165 t+0.09{\rm cos}(0.764t+3.6){\rm exp}(-0.06t)$ and $g(t)=0.24+0.0163t+0.11 {\rm cos}(0.765t+3.5){\rm exp}(-0.05t)$. The dashed lines $f_0(t)$ and $g_0(t)$ correspond to the $C_2$ rate for protocols A and B. The exact value of  $\frac{d}{dt}C_2$ for the IRLM at the self-dual point is $0.0165$ (see Sec.~\ref{sec:charge_fluct}).}
\label{fig:noise-time}
\end{figure}

We will also consider the second cumulant $C_2$ of the charge in one lead.
It is defined by
\begin{equation}
C_2(t)=\bra{\psi(t)} \hat Q^2 \ket{\psi(t)}  -\bra{\psi(t)} \hat Q \ket{\psi(t)}^2
\label{eq:C2def}
\end{equation}
where $\hat Q= \sum _{r=1}^{N/2}c_r^\dagger c_r$ is the operator measuring the total charge in the right lead.
A typical time evolution of this cumulant is presented in Fig.~\ref{fig:noise-time}. 
Since $C_2(t)$ grows linearly with time, a quantity of interest is the rate
\begin{equation}
S= \frac{d}{dt} C_2(t) \label{eq:S},
\end{equation}
which goes to a constant in the steady regime.
The long time limit of $S$~\footnote{Do not confuse with an entanglement entropy $S_{\rm vN}$.} is also a measure of the current {\em noise},
defined as the zero-frequency limit $\int_{-\infty}^\infty \langle\Delta \hat I (0) \Delta \hat I(\tau) \rangle d\tau$ of the current-current correlation function (see Apendix~\ref{sec:C2_S}).

This quantity will be studied in Sec.~\ref{sec:charge_fluct}, and its dependence on the bias will be compared with that of the BSG model.

\subsection{The BSG template}

Meanwhile, the current for the BSG model - after some simple manipulations to allow for a slight difference in geometry - admits two series expansions~\cite{fendley_exact_1995,fendley_exact_1995B}
 \begin{equation}
 I^{\rm BSG}={Vg\over 2\pi}\sum_{n=1}^\infty a_n(g)\left({V\over T_{\rm BSG}}\right)^{2n(g-1)}
 \label{eq:BSG_UV} 
  \end{equation}
at large $V$ (the UV regime) and 
\begin{equation}
I^{\rm BSG}={V\over 2\pi}-{V\over 2\pi g}\sum_{n=1}^\infty a_n(1/g)\left({V\over T_{\rm BSG}}\right)^{2n(-1+1/g)}. 
 \label{eq:BSG_IR}
\end{equation}
at small $V$ (the IR regime). Here $g={\beta^2\over 8\pi}$, 
 \begin{equation}
 a_n(g)=(-1)^{n+1} {\Gamma(3/2)\Gamma(ng)\over \Gamma(n)\Gamma(n(g-1)+3/2)}
 \end{equation}
and $T_{\rm BSG}\propto (\gamma')^{1/1-g}$.
 
It will be convenient in what follows to use the scaling form 
\begin{equation}
 I^{\rm BSG} = T_{\rm BSG} \;\vartheta(V/T_{\rm BSG})
\end{equation}
where e.g. at large $x$: 
%
%

\begin{equation}
\vartheta (x)={xg\over 2\pi}\sum_{n=1}^\infty a_n(g) x^{2n(g-1)}.
\label{eq:f}
\end{equation}

In the particular case $g=1/2$ the series can easily be summed to give
\begin{equation}
2\pi I^{\rm BSG}(g=1/2)={T_{\rm BSG}\over2}\hbox{arctan}{2V\over T_{\rm BSG}}
\end{equation}
This matches the well known result  for the RLM~\citep{bidzhiev_out-equilibrium_2017}  (IRLM at $U=0$) 
\begin{equation}
2\pi I^{\rm RLM}(U=0)=4t_B\hbox{arctan}{V\over 4t_B}
\end{equation}
after the identification $t_B={T_{\rm BSG}\over 8}$. Meanwhile, recall that if the perturbation in BSG is  normalized precisely as $2\gamma {\cos\beta\phi\over \sqrt{2\pi}}$  (so $\gamma'=\sqrt{2\over\pi}\gamma$) we have the relation
\begin{eqnarray}
T_{\rm BSG}=c_{\rm BSG}\gamma^{1/1-g}
\label{eq:TBSG2}
\end{eqnarray}
with
\begin{equation}
c_{\rm BSG}={2\over g}\left({\sqrt{2}\sin\pi g\Gamma(1-g)\over\sqrt{\pi}}\right)^{1/1-g}. 
\label{eq:c_BSG}
\end{equation}
So when $g={1\over 2}$, $T_{\rm BSG}=8\gamma^2$, and $t_B=\gamma^2$. 

\smallskip

We now propose to compare the measured $I$-$V$ curves for the IRLM in the scaling limit with the analytical expressions for the BSG current. In order to do this, we need to identify the coupling $g={\beta^2\over 8\pi}$ for the lattice IRLM: this can be done by studying
the algebraic decay of the current  in the large $V$ limit of the lattice model (see the next section for a precise definition)	
and comparing it with the prediction
for the continuum IRLM or BSG model. As will be discussed in Sec.~\ref{ssec:IR}, the later amounts to a perturbation theory of the 
of the continuum models by the boundary terms.
We are then left with the parameter $T_{\rm BSG}$ that we determine simply by a best fitting procedure. Note that, since the mapping on BSG is not supposed to work, there is no reason to use equation (\ref{eq:c_BSG}). We know by dimensional analysis that $T_{\rm BSG}\propto \gamma^{1/1-g}$, but it is interesting to see what a dependency of the prefactor on $g$ looks like, compared with (\ref{eq:c_BSG}).

\section{Current: numerics and comparison with the BSG results}
\label{sec:current}

\subsection{Power-law decay of the current at large bias, and associated exponent}
\label{ssec:large_bias}

\begin{figure}[h!]
\includegraphics[height=0.5\textwidth, angle =270]{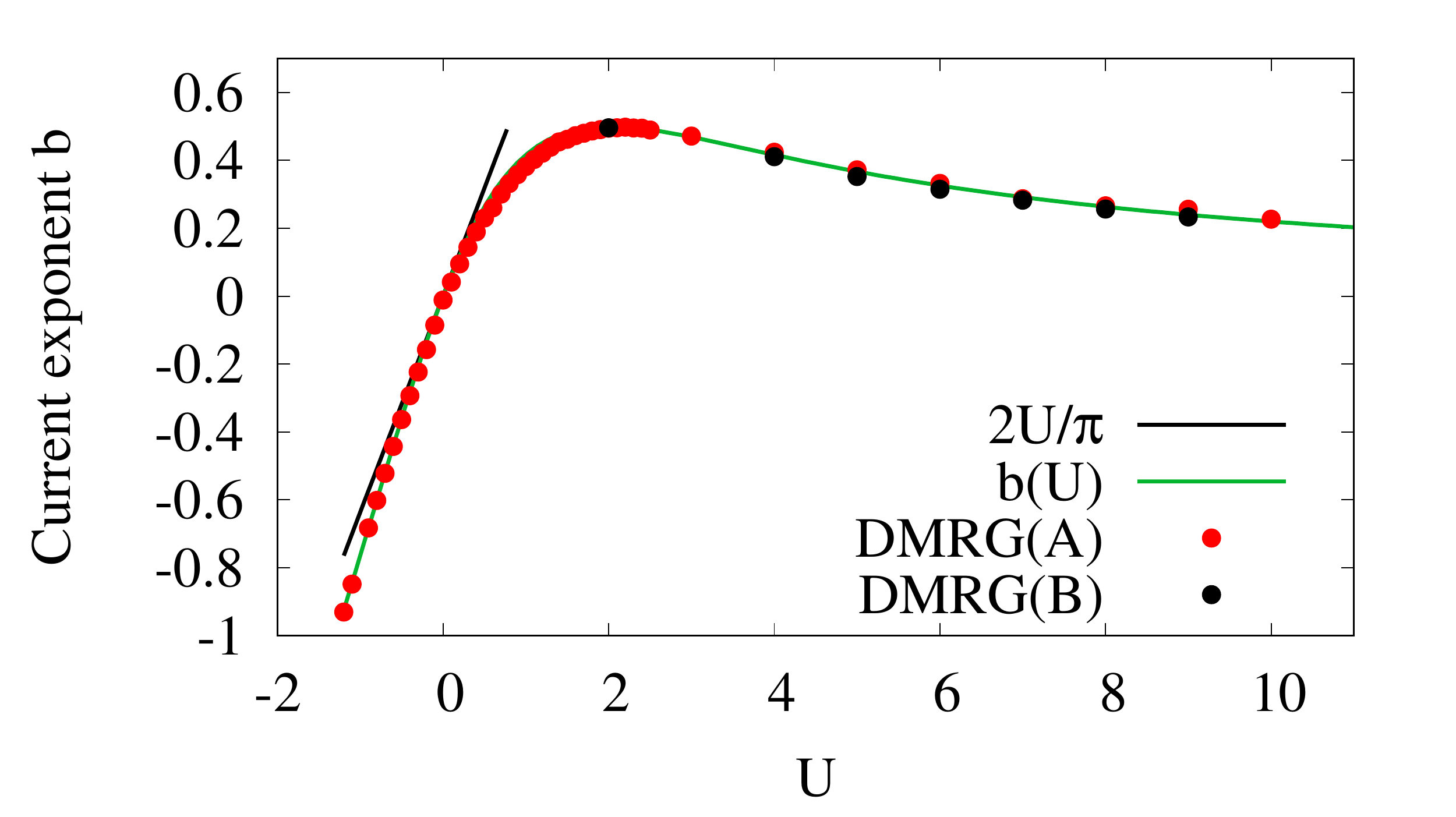}
\caption{Current exponent $b$ as a function of  $U$. Red (A) and black (B) dots were obtained by fitting the steady current to $I(V)=\text{cst}\cdot V^{-b}$. Green line: Eqs.~(\ref{eq:b})-(\ref{eq:U_c(U)}). Black line: perturbative expansion (small $U$)~\cite{karrasch_functional_2010,vinkler-aviv_thermal_2014}. Numerically, the maximal value of the exponent is $b_{\rm max}=0.494$, in good agreement with the exact value ($b_{\rm max}=1/2$).}
\label{fig:exp}
\end{figure}

In the scaling regime ($J'\ll J$ and $V\ll W$) the steady current of the IRLM vanishes as a power law
\begin{equation}
I_{\rm IRLM}(V) ={\rm cst}\cdot  V^{-b} \label{eq:def_b}
\end{equation}
in the large voltage limit~\cite{Doyon_exp,boulat_twofold_2008,bidzhiev_out-equilibrium_2017}, with an exponent given by
\begin{equation}
b=\frac{1}{2}\frac{U_c}{\pi}(2-\frac{U_c}{\pi}) \label{eq:b}.
\end{equation}
Here large $V$ means that $V$ is much larger than the scale $T_B$ (to be defined below, in Eq.~\eqref{eq:TB}) but still much smaller than the bandwidth $W$.
The interaction constant $U$ in the lattice model and its counter part $U_c$ in a continuum limit have a simple relation (see the supplemental material of Ref.~\onlinecite{FretonBoulat2014}):

\begin{equation}
U_c=\left\{
\begin{array}{cl}
4\arctan{(U/2)}, & {\rm }\;\; U < 2 \\
4\arctan{(2/U)}, & {\rm }\;\; U>2. \label{eq:U_c(U)}\\ 
\end{array}\right. 
\end{equation}
$b$ reaches the maximum $b_{\rm max}=\frac{1}{2}$  at the self-dual point located at $U=2$ (or equivalently $g=\frac{1}{4}$ and $U_c=\pi$), and
it is linear in $U$ close to $U=0$.
As shown in Fig.~\ref{fig:exp}, the exponent $b(U)$ extracted from the numerics by fitting the current in the large bias regime is in good agreement with the analytical formula [Eqs.~(\ref{eq:b})-(\ref{eq:U_c(U)})].
This is a quite nontrivial check of the validity and of the precision of the numerical procedure.
It should also be noted that for $U<0$  the exponent $b$ becomes negative, which means that the current keeps growing at large bias in presence of attractive interactions.
In what follows, the analysis of the numerical data will be made using the exact $b(U_c)$.

\begin{figure}[h]
\includegraphics[height=0.5\textwidth, angle =270]{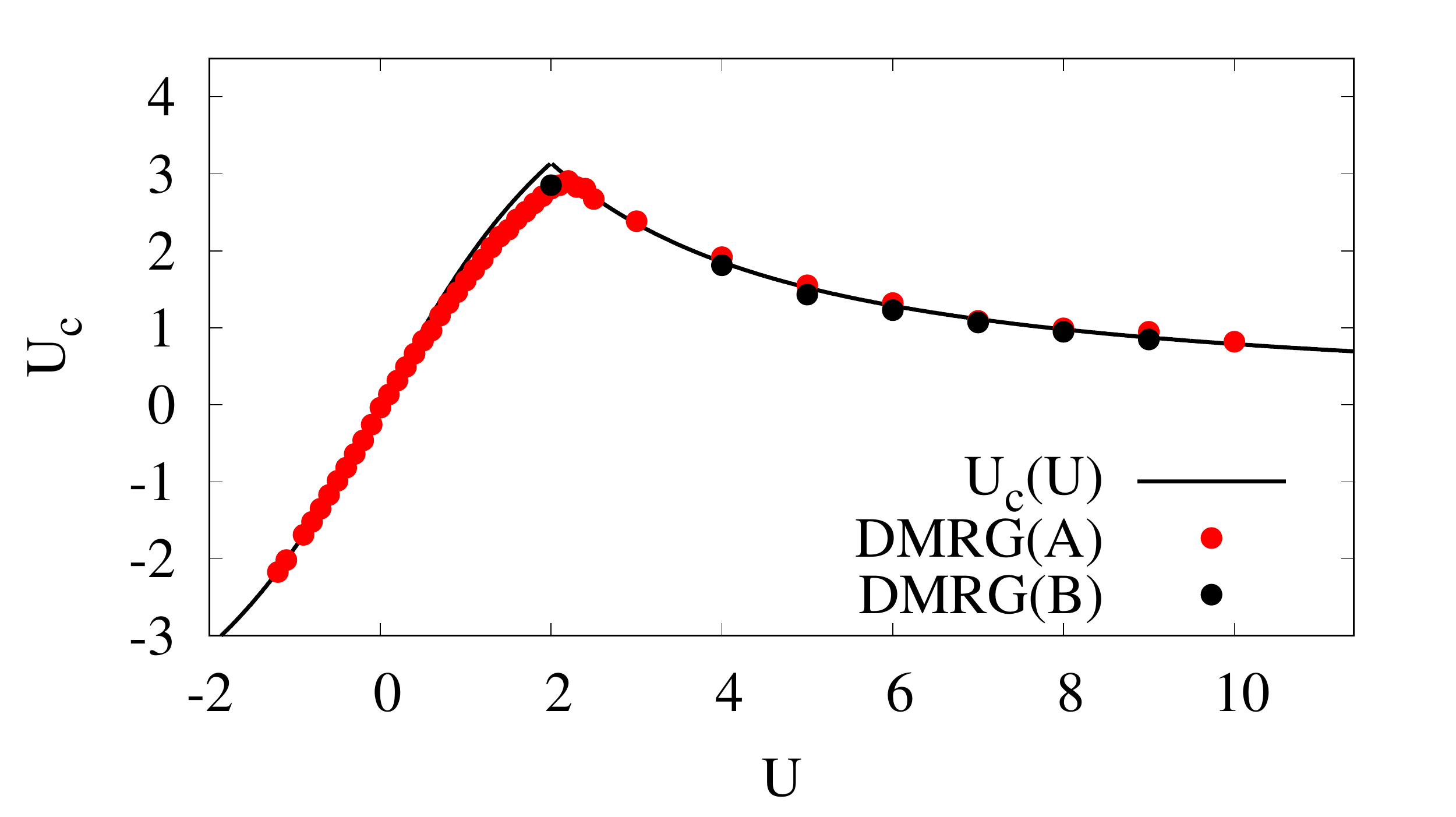}
\caption{Interaction strength in continuum limit $U_c$ versus interaction strength on the lattice $U$. Red (A) and black (B) dots are constructed by inverting Eq.~(\ref{eq:b}), where current exponent $b$ is obtained from tDMRG data; the black line represents Eq.~(\ref{eq:U_c(U)}).}
\label{fig:U}
\end{figure}

\subsection{Comparison with the $I$-$V$ curve of the BSG model}

The current-voltage curve of the IRLM is known exactly in two cases:  the noninteracting $U=0$ case \cite{branschadel_numerical_2010,landauer_spatial_1957,imry_conductance_1999,nazarov_quantum_2009,bidzhiev_out-equilibrium_2017} and self-dual point $U=2$~\citep{boulat_twofold_2008}. At these two points, the IRLM maps exactly to the BSG.
In this section we analyze to which extend the current $I_{\rm BSG}$ given by Eq.~(\ref{eq:BSG_UV}) could also describe
the current of the IRLM {\em away} from  the two cases above.
In other words, we will attempt to describe the current of the IRLM in terms of
the function $\vartheta$ [Eq.~(\ref{eq:f})] defined in Sec.~\ref{sec:IRLMvsBSG} for the BSG model.

\subsubsection{Large bias, $U$, $b$ and $g$}

The exponent $b$ of the IRLM is known exactly as a function of $U$ Eqs.~(\ref{eq:b}-\ref{eq:U_c(U)}).
In order to get the same large-bias exponent in the IRLM and in the BSG model,
the parameter $g$ of the BSG model has to become a function of $U$ (or $U_c$):
\begin{equation}
b=1-2g
\end{equation}
with $b$ given in Eq.~(\ref{eq:b}).

\subsubsection{Large bias and $c_{\rm IRLM}$ }
\label{ssec:cIRLM}

\begin{figure}[h]
\includegraphics[height=0.5\textwidth, angle =270]{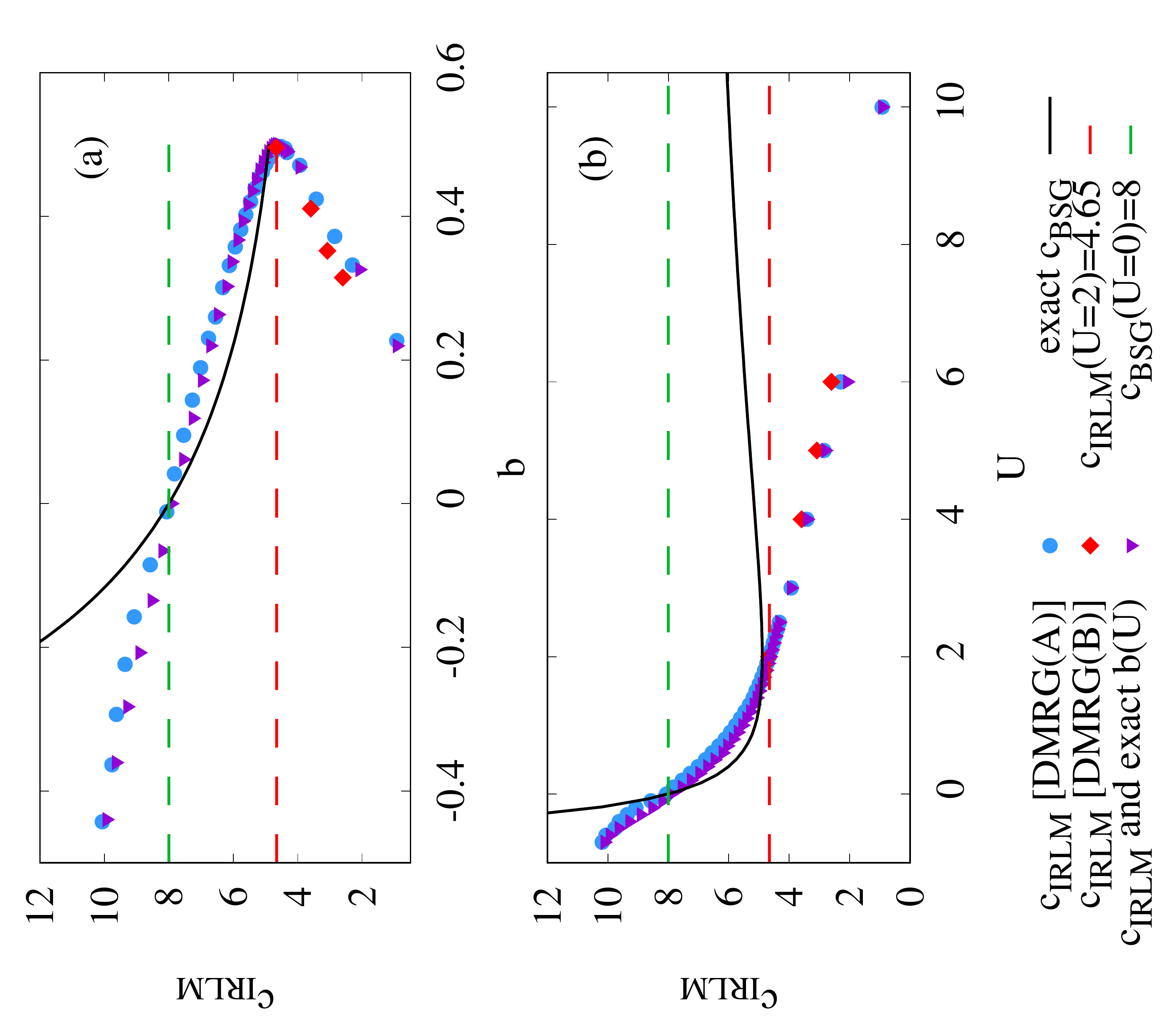}
\caption{Panel (a): coefficient $c_{\rm IRLM}$ as a function of $b$.
Blue dots and red diamonds: the exponent $b(U)$ and the coefficient $c_{\rm IRLM}(b)$ are obtained from the tDMRG numerics only for protocols A and B.
Purple triangles: the exponent $b(U)$ is taken from Eq.~(\ref{eq:b}), (\ref{eq:U_c(U)}), and $c_{\rm IRLM}(b)$ was extracted from the tDMRG data. Black line: $c_{\rm BSG}(b)$ given by Eq.~(\ref{eq:c_BSG}).
Red line: $2.7\frac{4\sqrt{\pi}\Gamma{(2/3)}}{\Gamma{(1/6)}}\approx 4.65$ (see~\cite{boulat_twofold_2008}) which corresponds to $U=2$ (self-dual point).
Green line: $c_{\rm BSG}(b=0)=8$ at $U=0$.
Panel (b): same but with $U$ on the horizontal axis.}
\label{fig:T_B}
\end{figure}

Since the prefactor $c_{\rm BSG}$ appearing in the definition of the energy scale $T_{\rm BSG}$ is {\it a priori} not a universal quantity,
it is natural to redefine it for the IRLM. In other words, to compare the current in the IRLM with that of the BSG model, we 
introduce a scale
\begin{equation}\label{eq:TB}
T_{\rm B}=c_{\rm IRLM} J'^{1/(1-g)},
\end{equation}
where $J'$ appears with the same {\em exponent} as $\gamma$ in $T_{\rm BSG}$ [Eq.~(\ref{eq:TBSG2})], but a different {\em prefactor}, $c_{\rm IRLM}$.
A related discussion on the prefactor of $(J')^{1/(1-g)}$ in the definition of $T_B$ can be found in a recent work by Camacho, Schmitteckert and Carr~\cite{camacho_exact_2019}.
In our case it is adjusted numerically (fit) so that the analytical curve for a given $g$ coincides with the tDMRG data at large bias:
\begin{eqnarray}
 I^{\rm IRLM} _{\rm numerics} \simeq   T_{\rm B} \; \vartheta( V / T_{\rm B}) \;\;{\rm when}\;V / T_{\rm B} \gg 1
 \label{eq:IRLM_vartheta}
\end{eqnarray}
[with $\vartheta$ defined in Eq.~(\ref{eq:f})].
The result of this procedure is a function $c_{\rm IRLM}(b)$ (top panel of
Fig.~\ref{fig:T_B}) or equivalently $c_{\rm IRLM}(U)$ (bottom panel of Fig.~\ref{fig:T_B}).
For comparison, we also plotted $c_{\rm BSG}$ Eq.~(\ref{eq:c_BSG}).

In case of the free-fermion problem, i.e. $g=1/2$ and $b=0$, the two models are equivalent and the theoretical value of crossover parameters is $c_{\rm BSG}=8=c_{\rm IRLM}$, in a good agreement with tDMRG data.
Since the IRLM at $U_c=\pi$ maps exactly onto the BSG model~\citep{boulat_twofold_2008}, we expect to have $c_{\rm IRLM}=c_{\rm BSG}$ at this point too.
The numerics give $c_{\rm IRLM}\approx4.66$ while the exact result is $c_{\rm BSG}=8\cdot \Gamma(3/4)^{4/3}/\pi^{2/3}\approx4.89$. This $5\%$ discrepancy is might be due to finite $J'$ effect, {\it i.e.} deviation from the scaling regime.

In Fig.~\ref{fig:T_B} we also marked the value of $c_{\rm IRLM}$ at the self-dual point which was found by Boulat {\it et~al.} \cite{boulat_twofold_2008}. Their estimate for $T_{\rm B}$ at this point is $2.7 c_0 (J')^{1/(1-g)}\approx4.65$ (with $g=1/4$ and 
$c_0=\frac{4\sqrt{\pi}\Gamma(2/3)}{\Gamma(1/6)}$). This value is in a good agreement with our data but differs by about $5\%$ from the exact value.
Away from the free-fermion point and away from the self-dual point, the curves for $c_{\rm IRLM}$ and $c_{\rm BSG}$ are significantly different.
$c_{\rm IRLM}$ monotonically decreases with increasing $U$ but $c_{\rm BSG}$ grows past the self-dual point at $U=2$. 
From this point of view, the models are thus generally not equivalent, as discussed in Sec.~\ref{sec:IRLMvsBSG}. As already mentioned,
the prefactors $c_{\rm IRLM}$ or $c_{\rm BSG}$ are not expected to be universal quantities, so the fact that $c_{\rm IRLM}\ne c_{\rm BSG}$ is not surprising at all.
We will now go further and
investigate if the  expression
of Eq.~(\ref{eq:IRLM_vartheta}) could also be used, at least approximately, for {\em finite} $V / T_{\rm B}$.

\subsubsection{Finite bias}

\begin{figure}[t]
\includegraphics[height=0.5\textwidth, angle =270]{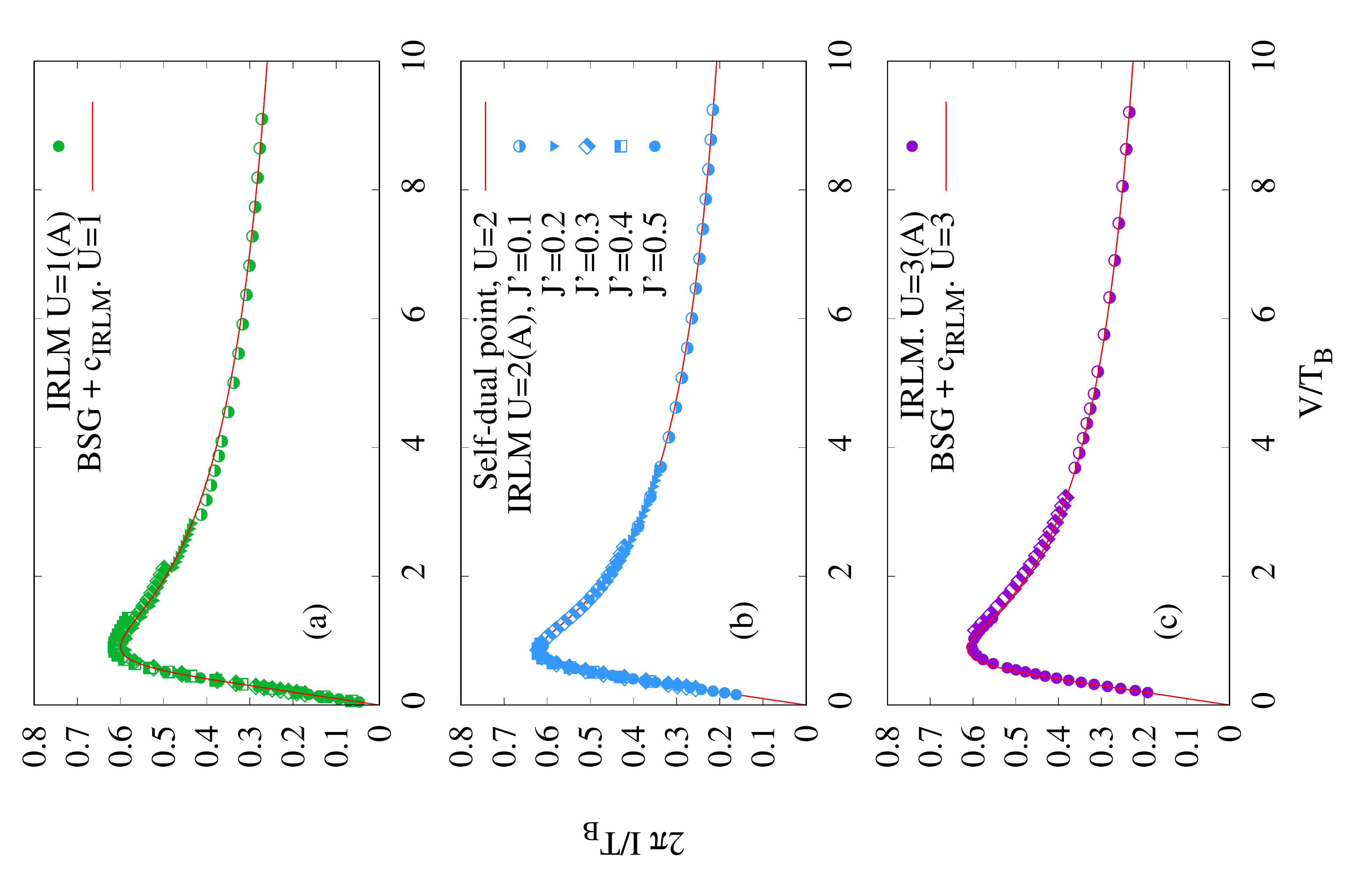}
\caption{Rescaled $I$-$V$ curves for different values of interaction, measured using the protocol (A). $U=1$ (a), $U=2$ (b) and  $U=3$ (c). The symbol shapes encode $J'$. As expected the agreement between the IRLM numerics and the BSG is excellent for $U=2$ (self-dual point), but it is
also very good for $U=1,3$ where the models are {\it a priori} {\em not} equivalent.}
\label{fig:Current}
\end{figure}

\begin{figure}[h!]
\includegraphics[height=0.5\textwidth, angle =270]{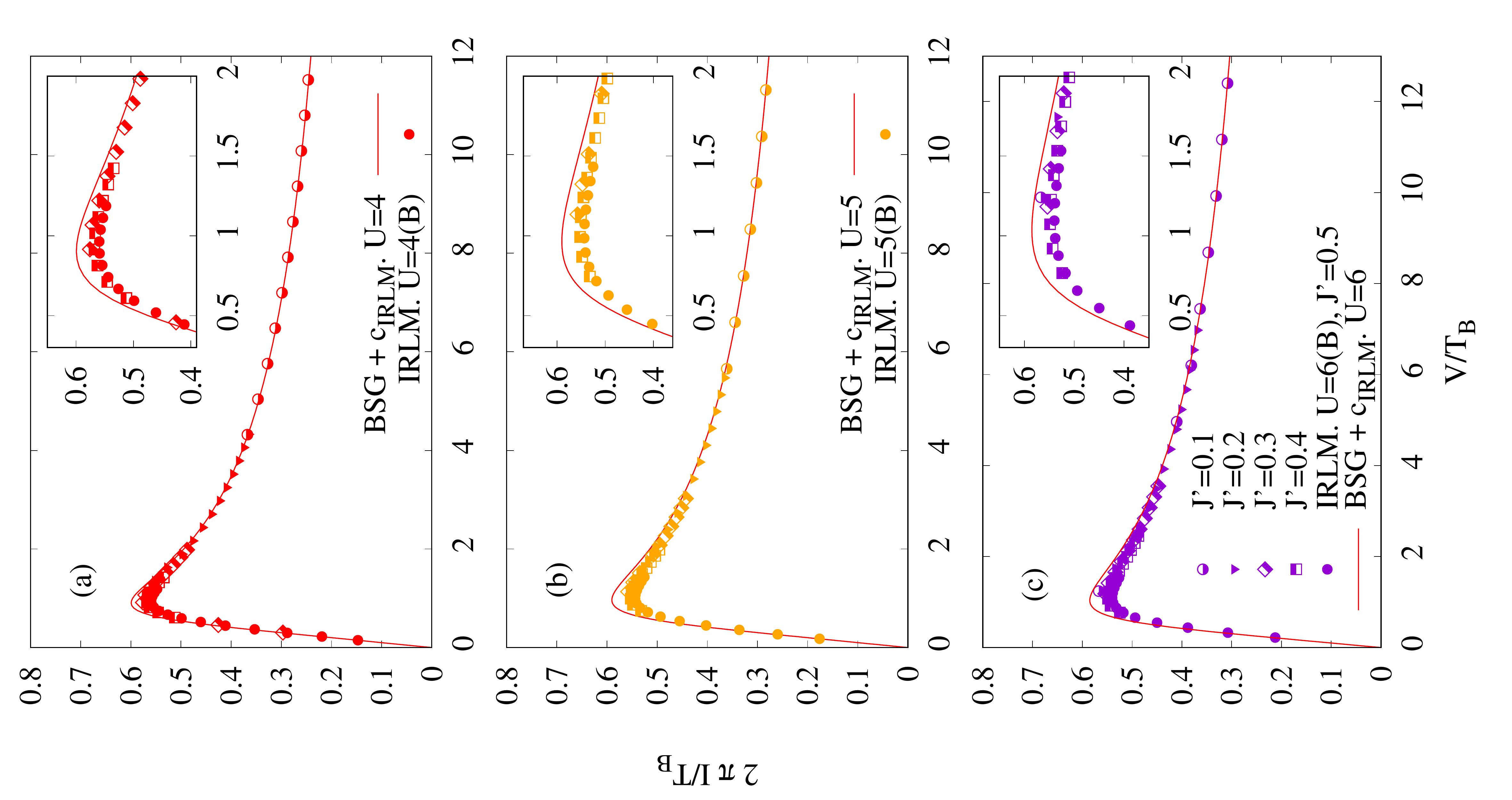}
\caption{
Same as Fig.~\ref{fig:Current} for $U=4$(a), $5$(b)  and $U=6$(c), using the quench protocol (B).
The data depart (insets) from the BSG curves [Eqs.~(\ref{eq:BSG_IR},\ref{eq:BSG_UV})] in the vicinity of $V/T_{\rm B}\simeq 1$.}
\label{fig:Cur_v2}
\end{figure}

Once the large-bias part of the current curve of the IRLM is adjusted to match that of the BSG (through $c_{\rm IRLM}$, as discussed
above), we can see if the agreement persists at lower bias. The results are displayed in Figs.~\ref{fig:Current}
and \ref{fig:Cur_v2}.

The remarkable and somewhat unexpected fact is that
for $U\lesssim 3$ 
the BSG function Eqs.~(\ref{eq:f})-(\ref{eq:IRLM_vartheta}) is a very good approximation of the IRLM current, even when $V/T_{\rm B}$ is of order 1.
While the agreement is excellent at the self-dual point (as it should, and as already noted in Refs.~\cite{boulat_twofold_2008,bidzhiev_out-equilibrium_2017,schwarz_nonequilibrium_2018})
the BSG function continues to describe  well the IRLM current {\em away} from $U=0$ and $U=2$. In fact, for $U\lesssim 3$, the deviation between BSG and IRLM is of the same order of magnitude as the numerical precision~\footnote{The latter precision can be estimated by looking at the approximate collapse of data obtained for different values of $J'$
(but same $V/T_{\rm B}$). The spread indicates to which extend the lattice model is close to the scaling limit. See also Appendix \ref{sec:numerics}.}
For $U$ from $4$ to $U=6$ (Fig.~\ref{fig:Cur_v2}) we start to observe some small discrepancy
between the numerical data and the BSG curves, close to the maximum of the current curve, when $V/T_B$ is close to 1.
The precise magnitude of this discrepancy is however
difficult to estimate since it is in this part of the $I$-$V$ curve the data sets associated with different values of $J'$ do not overlap perfectly.
Also, as can be seen in the insets of  Fig.~\ref{fig:Cur_v2}, the data points tend to get closer to the BSG curve when $J'$ is decreased.
So, the actual difference between the IRLM in the scaling regime and the BSG model might be smaller than what Fig.~\ref{fig:Cur_v2} indicates.

\begin{figure}[h!]
\includegraphics[height=0.5\textwidth, angle =270]{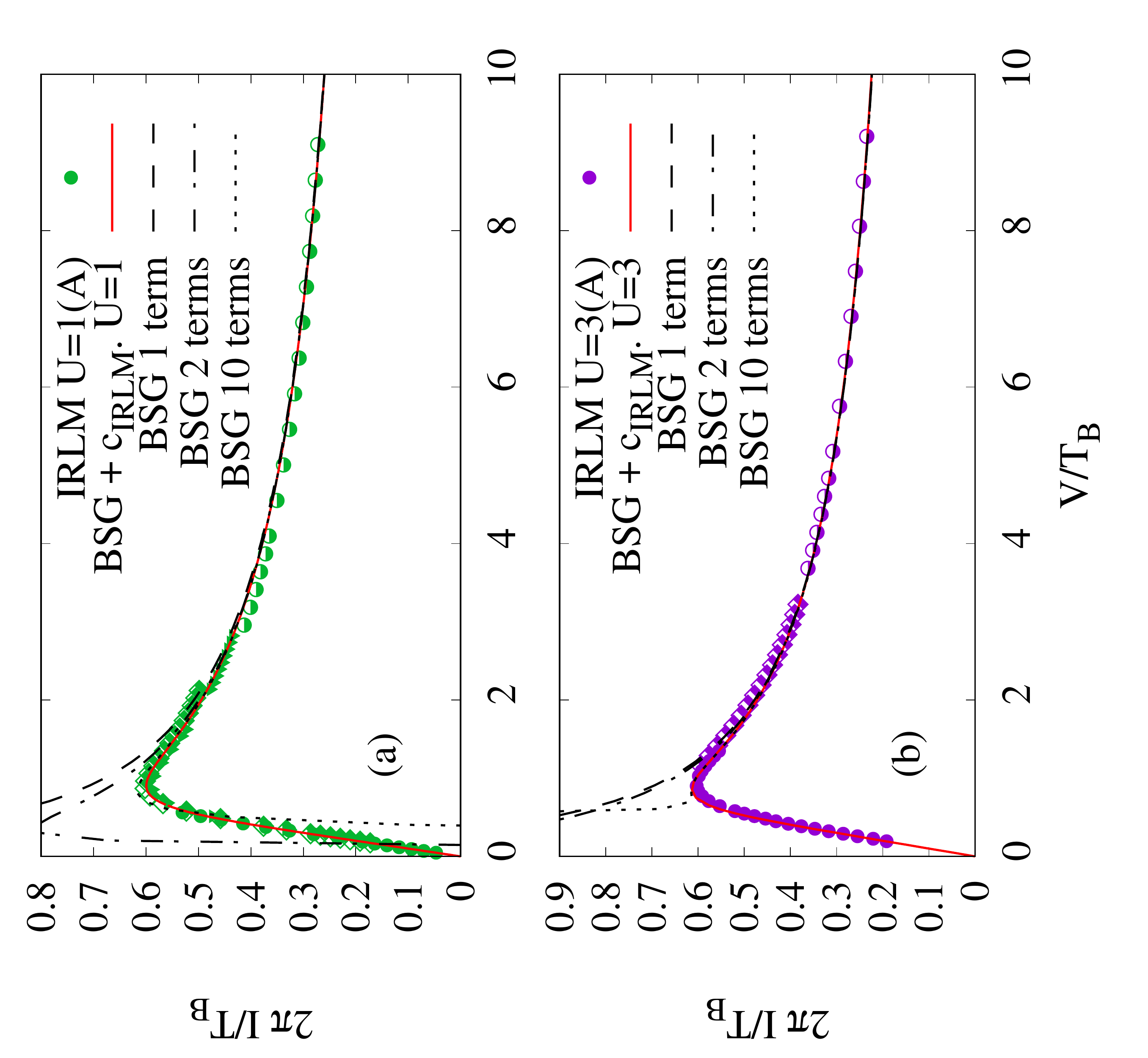}
\caption{Rescaled $I$-$V$ curves for the quench protocol (A) compared to different truncation orders in the large bias expansion
of the function $\vartheta$ [Eq.~(\ref{eq:f})] (black dashed and dotted lines). These expansions capture the behaviour if the IRLM in a wide range of rescaled bias $V/T_B$.}
\label{fig:terms}
\end{figure}

As commented earlier, it  is clear that one can expect a certain amount of similarity between the currents in the BSG and the IRLM. 
Thanks to our matching of the exponents and the $T_B$ scale (Sec.~\ref{ssec:cIRLM}), the leading terms
must agree by construction. On the other hand - as illustrated in  Fig.~\ref{fig:terms} which  shows (dotted or dashed lines) the leading term, or the sum of the first 2 or 10
terms in this expansion -  it is clear that the leading term only is not enough to reproduce the IRLM data close to the maximum of the current. The agreement between the $I$-$V$ curves for the two models in this region {\sl also} remains mysterious. What probably happens is that  the first few terms in the UV expansion are very close to each other. We have not been able to check this, because of the difficulty in calculating the higher-order terms in the Keldysh expansion of the IRLM current (this calculation is easier in the BSG model, in part because of the underlying integrability). But this raises the question: could it be that the field theoretic arguments in section \ref{sec:IRLMvsBSG} are flawed and that the two models out-of-equilibrium are in the same universality class? To answer this question, we turn again to the IR properties.

\section{Charge fluctuations and current noise}
\label{sec:charge_fluct}


To investigate further the differences between IRLM and BSG models, we consider the current noise,
as defined in Eq.~(\ref{eq:S}).
The results for the BSG model follow from more general calculations for the \textit{full counting statistics}~\cite{levitov_electron_1996} of BSG~\cite{fendley_exact_1995,fendley_exact_1995B}:
 \begin{equation}
 F^{\rm BSG}(\chi)={Vg\over \pi}\sum_{n=1}^\infty \frac{a_n(g)}{n} \left( {V\over T_{\rm BSG}} \right)^{2n(g-1)}\left( e^{i \chi n/2} -1\right)
 \label{eq:FCS_BSG_UV} 
  \end{equation}
where $\chi$ is a ``counting'' parameter. 
This also can be expanded at the low bias $V$ regime as
 \begin{eqnarray}
 F^{\rm BSG}(\chi)={V\over 2\pi} i \chi+{V\over \pi}\sum_{n=1}^\infty \frac{a_n(1/g)}{n} \times \nonumber\\ 
 \times \left( {V\over T_{\rm BSG}} \right)^{2n(-1+1/g)}\left( e^{-i \chi n/2g} -1\right).
 \label{eq:FCS_BSG_IR} 
\end{eqnarray}
From this function one can get the first charge cumulant, that is the mean current $I(V)=\partial F / i \partial \chi$ [Eqs.~(\ref{eq:BSG_UV}), (\ref{eq:BSG_IR})]. One can also get the current noise:
\begin{equation}
S(V)=-\frac{\partial^2 F}{\partial \chi ^2}.
\label{eq:C2}
\end{equation}
Note that charges have been normalized so that tunneling at the small coupling (large bias) is dominated by ${e\over 2}$ charges. (This convention corresponds to one electron tunneling from one wire to the other  in two steps). The expansion at low bias shows that, for BSG, the tunneling charge at large coupling is ${e\over 2g}$. 

The current noise has already been investigated numerically using tDMRG for the RLM ($U=0$)~\cite{branschadel_numerical_2010}
as well as at the self-dual point~\cite{branschadel_shot_2010,carr_full_2011}.
In Refs.~\cite{branschadel_numerical_2010,branschadel_shot_2010} $S$ was formulated in terms of the zero-frequency limit of the current-current correlations.
In Refs.~\cite{carr_full_2011,carr_full_2015}, using a modified time-evolution with an explicit counting field $\chi$, the cumulant generating function $F$ was estimated numerically
and the noise was extracted as the coefficient of the $\chi^2$ term.
More recently, a functional renormalization group approach~\cite{metzner_functional_2012} was used to 
compute the noise in the IRLM~\cite{suzuki_current_2016}\footnote{The functional renormalization group approach has also been employed to
study other aspects of the out-of-equilibrium Physics of the IRLM, see for instance Refs.~\cite{karrasch_functional_2010,kennes_renormalization_2012,kennes_interacting_2013}.},
especially in the regime of small $U$.
Instead, here we compute the current noise numerically using the relation between $S$ and the fluctuations of the charge,
as described by Eq.~(\ref{eq:S}). At any time, $C_2(t)$ is obtained by summing all the connected density-density correlations in the right lead:
\begin{eqnarray}
C_2(t)&=&\sum_{r,r'\geq 1} G(r,r') \\
G(r,r')&=&\langle \psi(t) |c^\dag_r c_r c^\dag_{r'} c_{r'} |\psi(t)\rangle\nonumber \\
&&-\langle \psi(t) |c^\dag_r c_r |\psi(t)\rangle\langle \psi(t) |c^\dag_{r'} c_{r'} |\psi(t)\rangle .
\end{eqnarray}
$S$ is then obtained by extracting (fits) the coefficient of the linear growth of $C_2(t)$ with time, as illustrated in Fig.~\ref{fig:noise-time}.

\begin{figure}[h!]
\includegraphics[height=0.5\textwidth, angle =270]{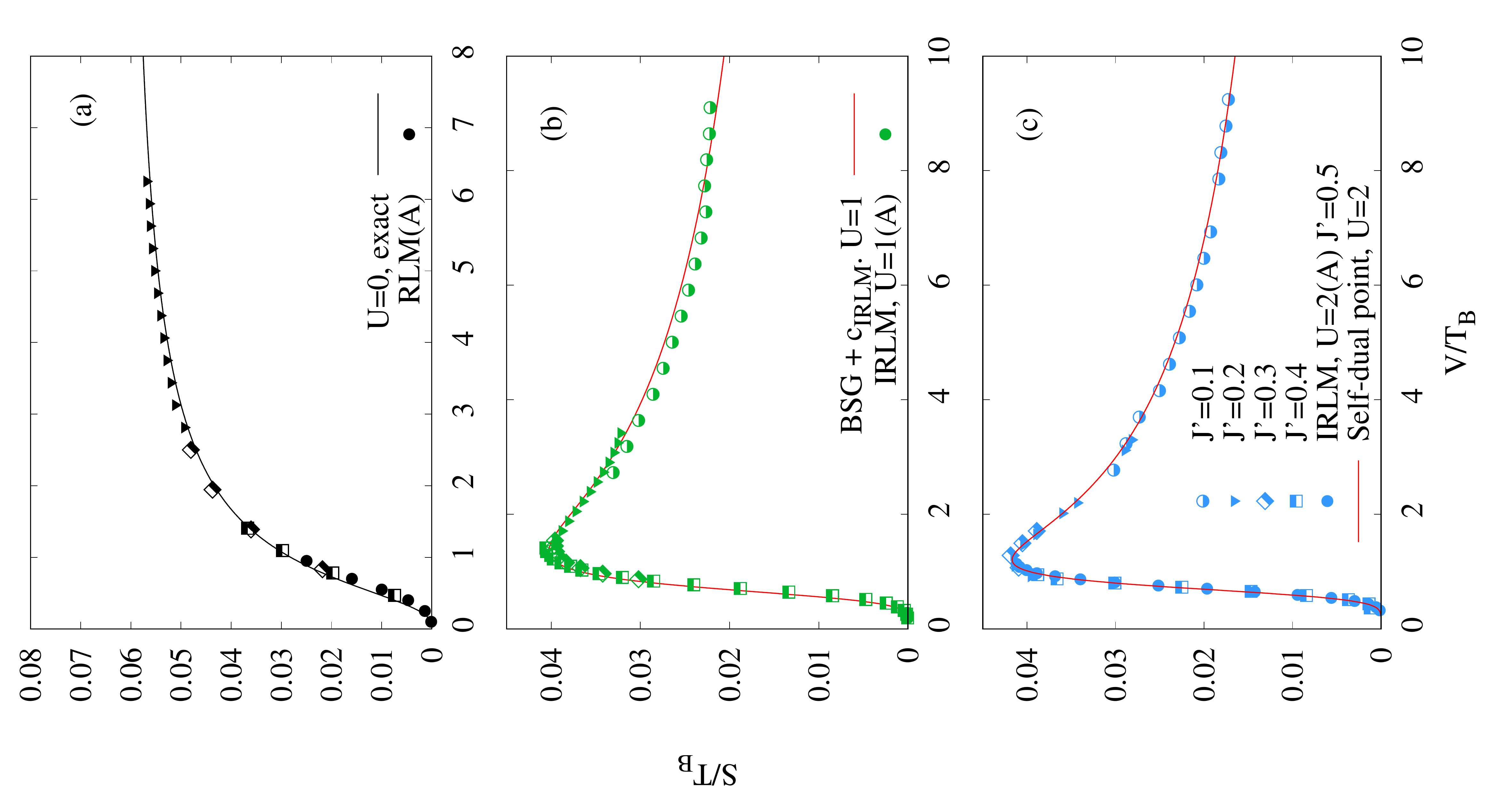}
\caption{Rescaled charge second cumulant rate $S/T_B$ versus $V/T_B$ for $U=0$(a), 1(b) and 2(c). The symbol shape encodes $J'$.
The full line for $U=0$ corresponds to Eq.~(\ref{eq:exactS_U=0}). For $U\ne0$ the full lines correspond to Eqs.~(\ref{eq:FCS_BSG_UV})-(\ref{eq:C2}).
Data obtained with the protocol A. These plots are constructed without any adjustable parameters: to obtain $T_B$, $b(U_c)$ is taken from Eqs.~(\ref{eq:b})-(\ref{eq:U_c(U)}) and $c_{ \rm IRLM}$ is determined from the large bias analysis of the current.}
\label{fig:noise}
\end{figure}

\begin{figure}[h!]
\includegraphics[height=0.5\textwidth, angle =270]{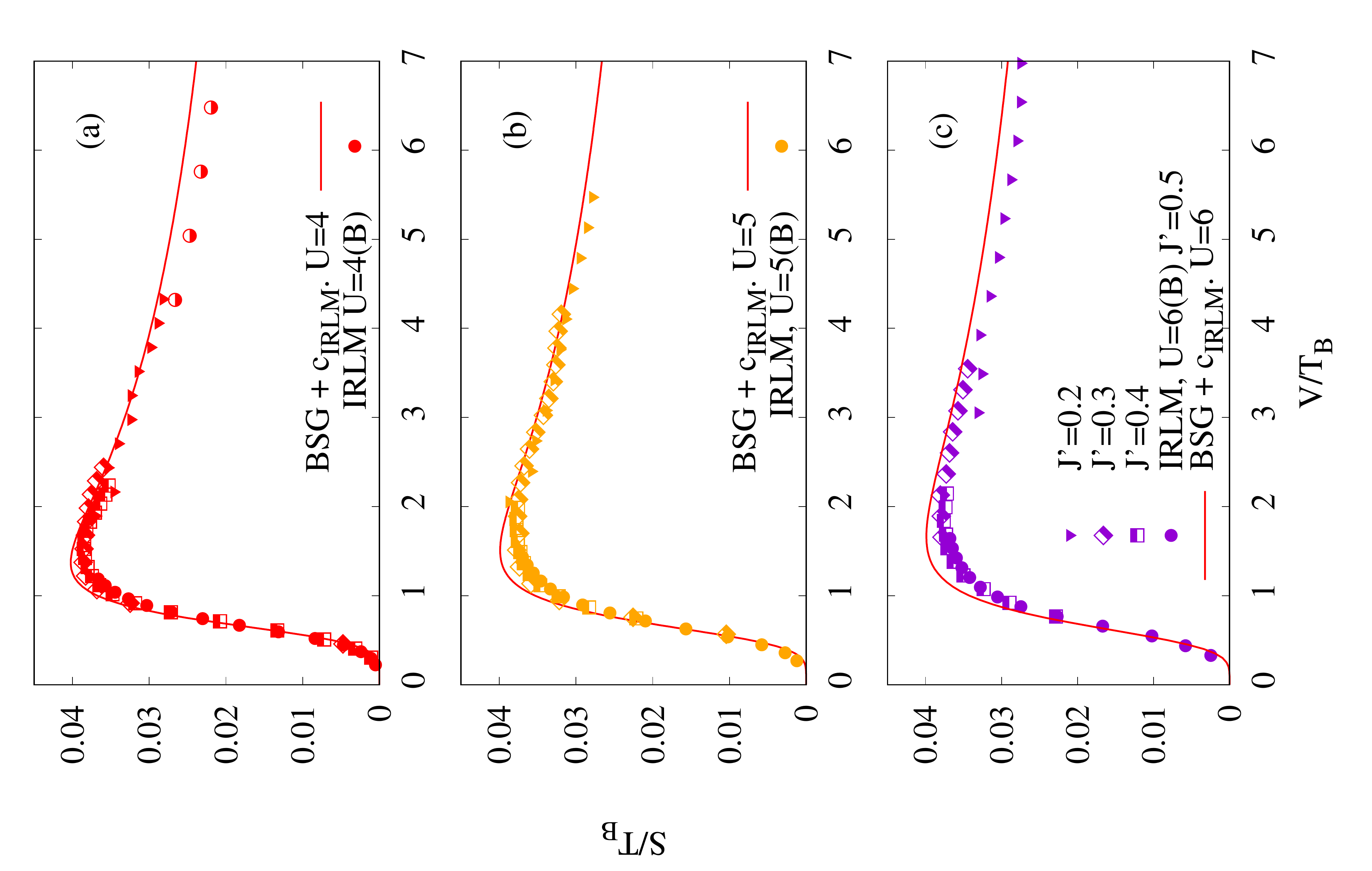}
\caption{Same as Fig.~\ref{fig:noise}, for $U=4$(a), 5(b) and 6(c) and the quench protocol (B).}
\label{fig:noise2}
\end{figure}

The resulting $S$-$V$ curves are presented in Figs.~\ref{fig:noise}-\ref{fig:noise2}
(see Fig.~\ref{fig:Noisebias} for some raw data, without rescaling). For $U=0$ (upper panel of Fig.~\ref{fig:noise}) the data are in good agreement with the exact free-fermion result: 
\begin{equation}
\frac{S(V)}{T_B}=\frac{1}{8\pi}\left(\arctan\left(\frac{2V}{T_B}\right)-\frac{2V/T_B}{1+\left(2V/T_B \right)^2}\right)
\label{eq:exactS_U=0}
\end{equation}
Our data at $U=0$ are also consistent with the results of Ref.~\onlinecite{branschadel_shot_2010}.

At $U=0$ and up to $U=2$ we observe a  good collapse of the rescaled curves (obtained for different values of $J'$) onto a single master curve, as for rescaled current $I(V)$ or the  entanglement entropy rate $\alpha(V)$~\cite{bidzhiev_out-equilibrium_2017}.
This indicates that, in the scaling regime, $S/T_{\rm B}$ is a function of the rescaled voltage $V/T_{\rm B}$.
Such a collapse remains visible for larger values of the interaction strength (Fig.~\ref{fig:noise2}), but we have a lower numerical precision at large $U$ and large $V/T_B$.

To analyze these data, we compare the shot noise $S/T_B$ of the IRLM with that of the BSG (full lines in Figs.~\ref{fig:noise}-\ref{fig:noise2}). We stress that there is
no new adjustable parameter, since for each $U$ and $J'$ we use the scale $T_B$ (and $c_{\rm IRLM}$) determined from the analysis of the current.
At $U=2$ (bottom panel in Fig.~\ref{fig:noise}) the data turn out to be in excellent agreement with the theoretical prediction for the BSG
[Eqs.~(\ref{eq:FCS_BSG_IR}), (\ref{eq:FCS_BSG_UV}) and (\ref{eq:C2})].
Quite remarkably, the agreement remains good away from the
two points where the models are known to be exactly equivalent, 
from $U=0$ up to $U\lesssim3$. This implies, in particular, that the shot noise in the IRLM decreases at large bias 
with an exponent which is very close (or identical) to that of the BSG model. The later is also the exponent
$b(U)$ describing the current decay in the large rescaled bias regime. The fact that the current and the shot noise
decay with the same exponent in the UV regime has already been noticed in \cite{suzuki_current_2016} for small (perturbative) $U$,
but it is observed here to hold at large $U$ too.

For very large values of $U$, above $\simeq 3$, 
some difference between the current noise of the IRLM and that of the BSG model
start to appear  (Fig.~\ref{fig:noise2}). So, with the precision of the present numerics, the shot noise in the IRLM and the BSG model can only be distinguished at large $U$  and in the crossover regime $V/T_{B}\simeq 1\cdots 2$.

\section{IRLM versus BSG}
\label{sec:IRLMvsBSG}

The agreement between the results for the IRLM and those for the BSG model is spectacular. Recall that the results for the two models are compared after two adjustments only: one for the anomalous dimension of the tunneling operator, and one for the crossover temperature. These two parameters are of course not enough to fix the whole crossover curves in general. In fact, were we not aware of the expected difference between the IRLM and the BSG model, a look at the curves would lead us to conclude that the two models are, in fact, ``probably in the same universality class" as far as their steady out-of-equilibrium properties are concerned. We do not think this is the case however, as we discuss now.

\subsection{The  field-theoretic description}

The IRLM  \cite{BoulatSaleur07}admits two field theoretic formulations which are close to, but in general not identical with, the BSG model. Both formulations are obtained using bosonization. Their difference originates in the fact  that one can first make linear combinations of the fermions in each lead then bosonize, or first bosonize, then make linear combinations of the resulting bosons. The first reformulation leads to  an anisotropic Kondo Hamiltonian:
\begin{equation}
H_{\rm IRLM}^{(1)}=\sum_{a=\pm} H_0(\phi_a)+{\gamma\over\sqrt{\pi}}\kappa_+ \left[e^{i\beta\phi_+(0)}S_++{\rm H.c}\right]\label{firstham}
\end{equation}
where $H_0(\phi_a)={1\over 2}\int dx (\partial_x\phi_a)^2$ is the free boson Hamiltonian.  Here the two leads have been unfolded so that the bosons are chiral, with equal-time commutators $[\phi_a(x),\phi_b(x')]=\delta_{ab}{i\over 4}\hbox{sign}(x-x')$.
We have $\phi_\pm={1\over\beta}\left[(\sqrt{4\pi}-\alpha)\varphi_\pm\mp\alpha\varphi_\mp\right]$, where $\alpha={U_c\over\sqrt{\pi}}$ and $\beta^2={2\over\pi}(U_c-\pi)^2+2\pi$; $\kappa_+=\eta\eta_+$. The fields $\varphi_\pm$ bosonize the even and odd combinations of {\sl physical fermions}
$\psi_\pm\equiv {1\over\sqrt{2}}(\psi_1\pm\psi_2)$
: $\psi_\pm={\eta_\pm\over\sqrt{2\pi}} e^{i\sqrt{4\pi}\varphi_{\pm}}$. Meanwhile,  $\eta_\pm$ are Klein factors.  The self-dual case is $U_{sd}=\pi$ corresponding to $\alpha=\sqrt{\pi}$, while the non-interacting case is $U_c=\alpha=0$. Finally, the amplitude $\gamma$ is related with amplitudes in the initial field theoretic formulation following \cite{BoulatSaleur07}. How this amplitude is related with the tunneling term in the lattice Hamiltonian (the ``bare'' coupling constant) will be discussed below. 

The second reformulation mixes somehow the Kondo and the BSG Hamiltonians, and reads
\begin{equation}
H_{\rm IRLM}^{(2)}=\sum_{a=\pm} H_0(\phi_a)+{\gamma\over\sqrt{2\pi}} \left[V_1(0)O_2(0)S_++{\rm H.c}\right]\label{secondham}
\end{equation}
where $V_{\pm 1}=e^{\pm i\beta_1\phi_1},~O_2=\kappa_1 V_2+\kappa_2V_{-2}$, $V_{\pm 2}=e^{\pm i\sqrt{2\pi}\phi_2}$, $\kappa_a=\kappa\eta_a$. We have set  $\beta_1=\sqrt{2\pi}-\alpha\sqrt{2}$, so $\beta^2=\beta_1^2+2\pi$.
The fields are now
 $\phi_1={1\over \sqrt{2}} (\varphi_1+\varphi_2)$ and $
\phi_2={1\over \sqrt{2}} (\varphi_1-\varphi_2)$ where $\varphi_{1,2}$ bosonize the fermions $\psi_{1,2}$ in each lead independently.

The voltage difference between the leads translates into a  term $V$ coupled to the charge $\int \partial_x\phi_2$.  The current can be calculated using the Keldysh method and the Hamiltonian (\ref{secondham}) where $V_{\pm 2}\to e^{\pm iVt}V_{\pm 2}$. 
\bigskip

We now turn to the   BSG model 
\begin{equation}
H_{\rm BSG}=H_0(\phi_2)+\gamma' \cos\beta\phi_2(0)\label{BSGham}
\end{equation}
with Hamiltonian chosen in such a way that the tunneling term has the same dimension $g\equiv{\beta^2\over 8\pi}$ as in the IRLM, and $\gamma'\propto \gamma$. 
The voltage can similarly be introduced by $e^{\pm i\beta\phi_2}\to e^{\pm i(\beta\phi_2+Vt)}$. The difference between (\ref{secondham}) and (\ref{BSGham}) is obvious: the second Hamiltonian involves only two vertex operators, and does not contain any spin. This leads to profoundly different RG flows in equilibrium: for the BSG model, the flow is between Neumann  and Dirichlet boundary conditions, with, for instance, a ratio of boundary degeneracies \cite{AffleckLudwig,FendleySaleurWarner} that depends on $\beta$. For the IRLM the flow is between Neumann with a decoupled extra spin and Neumann, and the ratio of boundary degeneracies is equal to $2$, irrespective of the coupling $\beta$. 

\subsection{IR physics}
\label{ssec:IR}

The  Hamiltonian (\ref{BSGham}) definitely looks different from  (\ref{firstham}, \ref{secondham}) - and, in particular, does not involve a discrete spin degree of freedom. As just discussed, some of their equilibrium properties definitely {\sl are} different. This is however no proof  at all that the corresponding steady-state out-of-equilibrium properties are not in the same universality class. Certainly, as far as transport is concerned, the UV and IR fixed points are  similar in both models. Moreover, expansions in powers of the UV coupling constants ($\gamma',\gamma$) are also formally identical: if  say the $T=0$ $I$-$V$ curves are different in both models, this can only be because of {\sl quantitative} differences in these expansions. Unfortunately, since we compare the two models after adjusting the crossover scale, such quantitative difference can only be seen after a sixth order Keldysh calculation of the current for the IRLM in powers of $\gamma$ (the corresponding expansion  for the BSG model is 
known, thanks in part to its integrability). The required calculations are not present in the literature, and are beyond the scope of this work. 

\smallskip

Luckily, we can also explore the potential difference between  transport in the IRLM and BSG model by considering instead the vicinity of the  IR fixed point. This is the limit of close to  perfect transmission (with, for the IRLM, the impurity spin hybridized with the leads), where the voltage is small, or the coupling constants $\gamma,\gamma'$  large. Of course, this regime  is usually very difficult to control since it it {\sl outside  the perturbative domain}.  In our case however, the difficulty can be bypassed  using general techniques of integrability {\sl in equilibrium}.

\smallskip

To see what happens, it is more convenient  to use the Hamiltonian (\ref{firstham}). According to \cite{LesageSaleur99}, the approach to the IR fixed point is given by an infinite series of operators ${\cal O}_{2n}$ which are well defined expressions in terms of the stress energy tensor, and whose coefficients are known exactly, and scale as $\gamma^{-(2n-1)/1-g}$, where $g$ is the dimension  of the perturbation in the UV, $g={\beta^2\over 8\pi}$. We have for instance~\footnote{Note: from now on, we suppress mention of the coordinate in the exponentials of the fields at the origin. We do instead mention the time coordinate whenever we use a Heisenberg representation.}

\begin{eqnarray}
{\cal O}_2&=&{1\over 2\pi}\ T\nonumber\\
{\cal O}_4&=&{1\over 2\pi}:T^2:\nonumber\\
{\cal O}_6&=&{1\over 2\pi}\left(:T^3:-{c+2\over 12}:T\partial^2T:\right)
\end{eqnarray}
while
\begin{equation}
T=-2\pi :(\partial\phi_+)^2:+i(1-g)\sqrt{{2\pi\over g}} \partial^2\phi_+\label{stresstens}
\end{equation}
and $c=1-6{(1-g)^2\over g}$. Here all the fields are defined exactly like before. 

To understand the argument, it is now enough to consider the derivatives 
\begin{eqnarray}
\partial\varphi_+\sim (\psi_1^\dagger+\psi_2^\dagger)(\psi_1+\psi_2)\sim \partial\phi_1+\cos\sqrt{8\pi}\phi_2\nonumber\\
\partial\varphi_-\sim (\psi_1^\dagger-\psi_2^\dagger)(\psi_1-\psi_2)\sim \partial\phi_1-\cos\sqrt{8\pi}\phi_2
\end{eqnarray}
where $\sim$ means up to proportionality coefficients in all the terms on the right hand side. In the generic case, it follows that $\partial\phi_+$ is a linear combination of $\partial\phi_1$ and $\cos\sqrt{8\pi}\phi_2$. Therefore, the stress tensor term $T$ (\ref{stresstens}) is a combination involving, as far as charge transferring terms are concerned,  $\partial\phi_1\cos\sqrt{8\pi}\phi_2$ and $\partial\phi_2\sin\sqrt{8\pi}\phi_2$ -  the latter term coming from the $\partial^2\phi_+$. Now - and this is the crucial point - when considering $:T^2:$ and the products
$$
\partial\phi_{1({\rm resp.} 2)}(z)\cos\sqrt{8\pi}\phi_2(z)\partial\phi_{1({\rm resp.} 2)}(w)\cos\sqrt{8\pi}\phi_2(w)
$$
and using 
$$
\partial\phi_{1({\rm resp.}2)}(z)\partial\phi_{1({\rm resp.}2)}(w)\sim {1\over (z-w)^2}+\ldots
$$
together with
\begin{eqnarray}
\cos\sqrt{8\pi}\phi_2(z)\cos\sqrt{8\pi}\phi_2(w)\sim \nonumber \\ 
{1\over (z-w)^2}(1+\ldots)+(z-w)^2(\cos2\sqrt{8\pi}\phi_2(w)+\ldots)
\end{eqnarray}
we will generate in $:T^2:$ a term $\cos2\sqrt{8\pi}\phi_2$. Meanwhile, the bosonized expression of  $T$ itself involved a  $\partial\phi_1\cos\sqrt{8\pi}\phi_2$ term, so we see we generate terms corresponding to different transfers of charge - that is, different integer multiples of  $\sqrt{8\pi}\phi_2$ in the exponentials. Meanwhile, all these terms come with the proper power of the UV coupling constant $\gamma$, in this case:
\begin{eqnarray}
\partial\phi_1\cos\sqrt{8\pi}\phi_2,~~\hbox{coupling}~~\gamma^{-1/(1-g)},~~\hbox{charge}~~e\nonumber\\
\cos2\sqrt{8\pi}\phi_2,~~\hbox{coupling}~~\gamma^{-3/(1-g)},~~\hbox{charge}~~2e
\end{eqnarray}
We see that the transfer of charge in the IR involves in general two terms at leading order,  transporting respectively $e$ and $2e$, with different scaling coefficients. The full counting statistics~\cite{levitov_electron_1996} at leading order for instance would be a function of $\gamma^{-1/(1-g)}e^{-i\chi}$ and $\gamma^{-3/(1-g)}e^{-2i\chi}$ (where $\chi$ is the counting variable coupled to the charge). In particular, the leading term always corresponds to tunneling of electrons. This is in agreement with the fact that the anisotropic Kondo fixed point is a Fermi liquid. 

Note that the argument breaks down if the amplitude of the leading term $\gamma^{-1/(1-g)}e^{-i\chi}$ happens to vanish.  This is precisely what happens in the self-dual case, since then $\partial\phi_+$ contains the term $\cos\sqrt{8\pi}\phi_2$ only. In that case, the bosonized version of $T$ does not contain a $\cos\sqrt{8\pi}\phi_2$ term anymore, while the $:T^2:$ term still contains a $\cos2\sqrt{8\pi}\phi_2$ term as before. In fact, one can show that at all orders in the ${\cal O}_{2n}$, all that appears are $\cos2\sqrt{8\pi}\phi_2$ terms. Hence the transferred charges in this case are multiples of $2e$, not of $e$ - a fact   in agreement with the equivalence with the boundary sine-Gordon model at this point. Note that the result  is not incompatible with the fixed point being Fermi liquid: what happens is simply that amplitudes in the mapping conspire to cancel the usual term describing tunneling of single electrons, and what is observed is tunneling of pairs instead.

The other case where the argument breaks down is the free-fermion case $g={1\over 2}$. In this case indeed, all quantities ${\cal O}_{2n}$ can be expressed solely as 
\begin{equation}
{\cal O}_{2n}\sim \psi_+^\dagger\partial_{2n-1}\psi_+
\end{equation}
and thus all involve only $\cos\sqrt{8\pi}\phi_2$, corresponding to the transfer of charges $e$, and an expansion for the FCS in terms of $\gamma^{-2}e^{-i\chi}$. 

\smallskip
To summarize this technical section: tunneling  in the IR for the IRLM  involves in general  $\gamma^{-1/(1-g)}e^{-i\chi}$ and $\gamma^{-3/(1-g)}e^{-2i\chi}$, corresponding at leading order to transfers of charge $e$ and charge $2e$. It involves the first combination only when $g={1\over 2}$ and the second only when $g={1\over 4}$. The IR behavior is thus dominated by transfer of electrons for all values of $U$ but $U_{sd}$, where it is then dominated by transfer of {\sl pairs} of electrons instead. This extends a result of Ref.~\cite{suzuki_current_2016}, which stated that the effective charge is $e\left[1+\mathcal{O}(U^2)\right]$, and which was therefore relevant only in the vicinity of the noninteracting RLM.
\smallskip

\begin{figure}[h!]
\includegraphics[height=0.5\textwidth, angle =270]{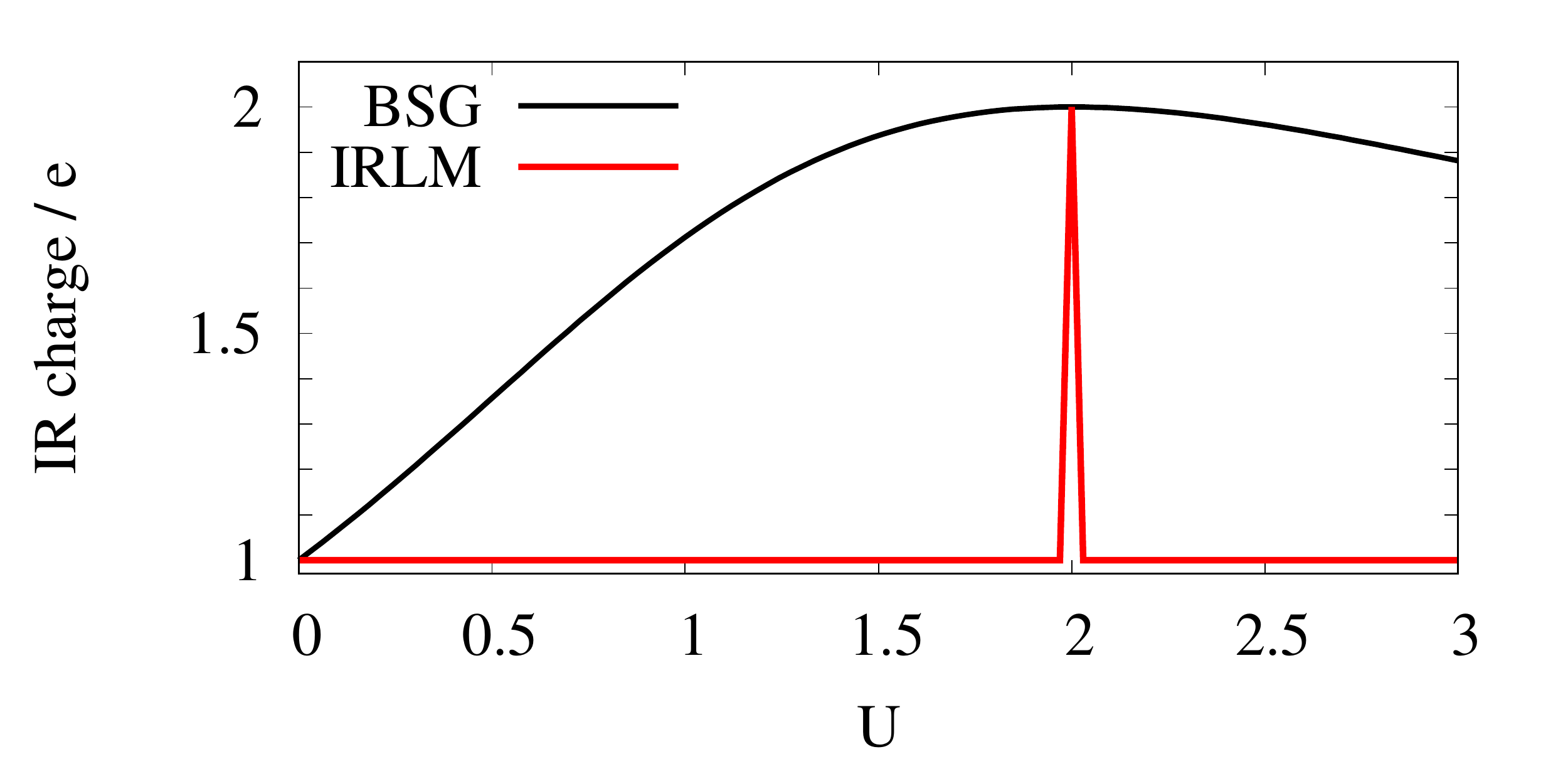}
\caption{Charge of quasiparticles dominating transport in the deep IR  for  the IRLM and BSG model. The charge for the  IRLM differs from $e$ only at the self-dual point where it becomes equal to $2e$, whereas the charge for the BSG model  is a  nontrivial function of the interaction parameter $U$.}\label{fig:IR_q}
\end{figure}

This fact is the best way to state physically the difference between the IRLM and the BSG models. For the BSG model (\ref{BSGham}) the approach to the IR fixed point is described, up to operators that do not transfer charge, only by the operator $\cos{8\pi\over\beta}\phi_2$. This means that, if the tunneling charge in the UV is normalized to be ${e\over 2}$, the charge in the IR is ${4\pi\over \beta^2}e$. This is $2e$ at the self-dual point, $e$ at the non-interacting point, and a {\sl non-trivial, coupling dependent, not integer multiple of $e$ in between} (Fig.~\ref{fig:IR_q}). Hence the nature of charge transfer in the two models should be profoundly different in the infrared. This can be seen explicitly in the behavior of two different quantities: the small bias backscattered current, and the shot noise. 

\subsection{Small bias and backscattered current}
\label{ssec:small}

The difference in the value of tunneling charges has consequences on the $I-V$ characteristics. In fact,  
the low-bias expansion of the steady current in the IRLM has been computed up to order $\mathcal{O}(V^6)$
by Freton and Boulat~\cite{FretonBoulat2014}. They computed the backscattered current $I_{\rm BS}=V/2\pi-I$, {\it ie} the difference between the current $I$ and the value
of the current in absence of impurity. Their result reads~\footnote{
For the Freton-Boulat expansion to match the exact formula at the free-Fermion point,
we had to change the sign of the third term in their formula.}:
\begin{eqnarray}
I_{\rm BS}&=&\frac{XV^3}{48g^2 T_B^2} \left[ \begin{array}{c} \\ \\ \end{array}
\right.\nonumber \\
&& 1+ \frac{3\kappa_4*3V^2\left(X^2-10X+5\right)}{40gT_B^2}  \nonumber \\
&&\left. +\frac{3V^2\left(X^2+1 \right)}{40g^2T_B^2} \right]+\mathcal{O}\left(T_B^{-6}\right),
\end{eqnarray}
where $X=4g-1$ and 
\begin{equation}
\kappa_{2n}=\frac{(g/\pi)^{n-1}}{(n-\frac{1}{2})n!}\frac{\Gamma \left( \frac{2n-1}{2\left(1-g \right)} \right)}{\Gamma \left( \frac{g\left(2n-1 \right)}{2\left(1-g \right)} \right)} \left[ \frac{\Gamma \left( \frac{g}{2\left(1-g \right)} \right)}{\Gamma \left( \frac{1}{2\left(1-g \right)} \right)}\right]^{2n-1}.
\end{equation}

\begin{figure}[h!]
\includegraphics[height=0.5\textwidth, angle =270]{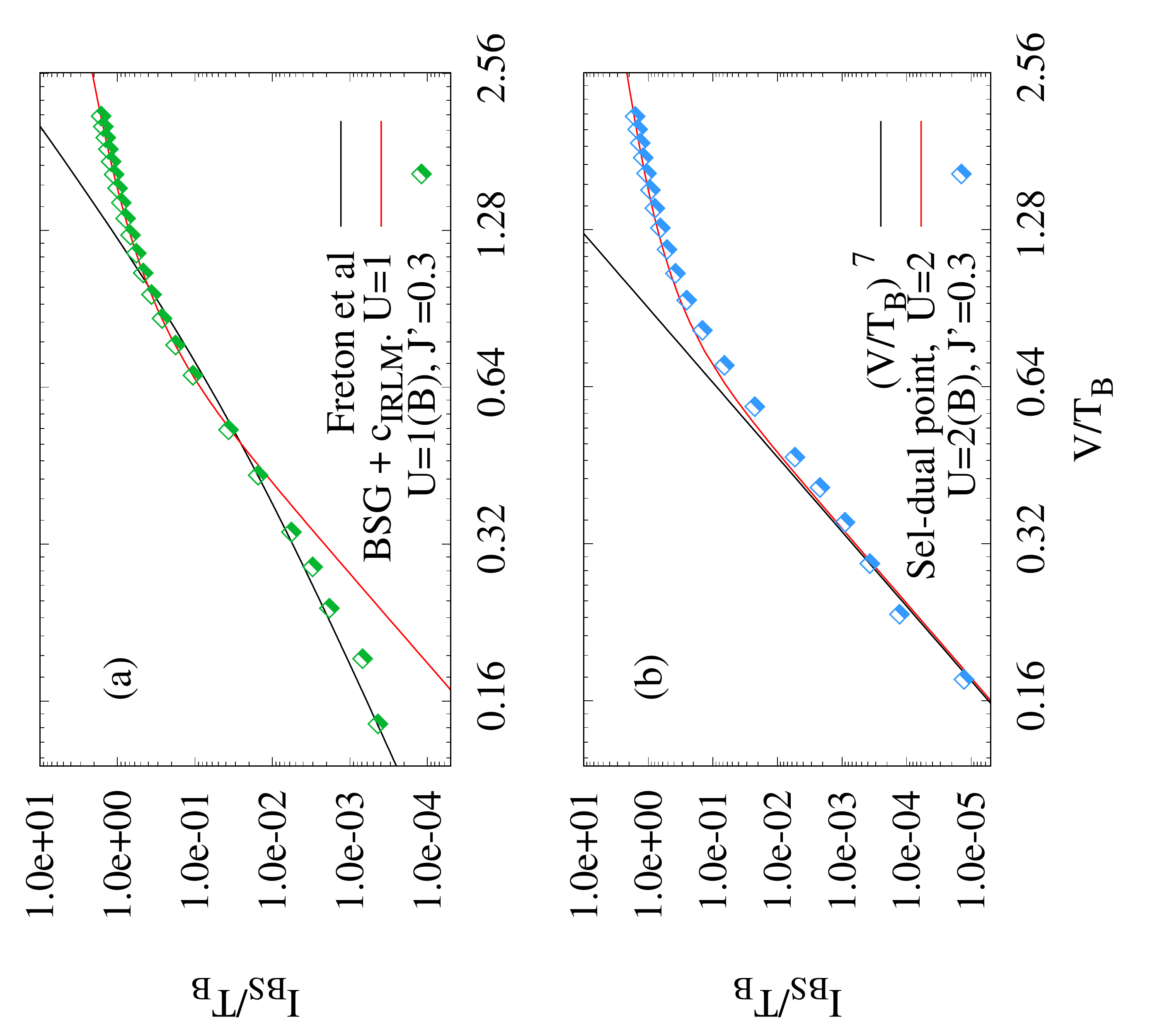}
\caption{Panel (a): Rescaled backscattered current $I_{\rm BS}/T_{\rm B}=V/(2 \pi T_{\rm B})- I/T_{\rm B}$ as a function of $V/T_{\rm BS}$ at $U=1$.
Panel (b): same for $U=2$ (self-dual point). For $U=2$, $I_{\rm BS}$ vanishes as $~V^7$
at low $V$, as expected from the equivalence with the BSG at the self-dual point.
For $U=1$, the data are consistent with $~V^3$, as expected for the IRLM
away from the self-dual point.
Data obtained using the protocol B.}\label{fig:IBS}
\end{figure}

This shows that for the IRLM $I_{\rm BS}$ vanishes as $V^3$ at low bias.
On the other hand, the backscattered current in the BSG model can be read of Eq.~(\ref{eq:BSG_IR}),
and its leading term has an exponent which varies continuously with $g$:
$I_{\rm BS}^{\rm BSG} /T_{\rm BSG} \sim \left(V/T_{\rm BSG}\right)^{2/g-1}$. So, for sufficiently low $V$ the numerical data
for the IRLM should show a $V^3$ behavior and should depart from the BSG results if $U\ne 0$ and if $U\ne 2$. At the self-dual point the coefficient of the $V^3$ and $V^5$ terms vanish and
the leading term in $I_{\rm BS}$ becomes $\mathcal{O}(V^7)$ \cite{FretonBoulat2014}, in agreement with the BSG result at $g=1/4$. All these features can be understood with an IR perturbative analysis similar to the one sketched in section \ref{sec:IRLMvsBSG}.
The data plotted in Fig.~\ref{fig:IBS} illustrate the  low-bias behaviors of $I_{\rm BS}$ at the self-dual point (bottom panel), and at a more generic value of $U$ (upper panel). We see that, in fact, the currents in the BSG and IRLM have different analytical behaviors in this region - and that these properties are in agreement with the field theoretic analysis.
While the low-bias data at $U=1$ follow the perturbative
prediction for the IRLM as it should, it is striking that $I_{\rm BS}$ then joins the BSG curve for $V/T_B \gtrsim 0.5$,
although there is a priori no reason why it should do so at such intermediate bias.

\subsection{Charge of the carriers}

\begin{figure}[h!]
\includegraphics[height=0.5\textwidth, angle =270]{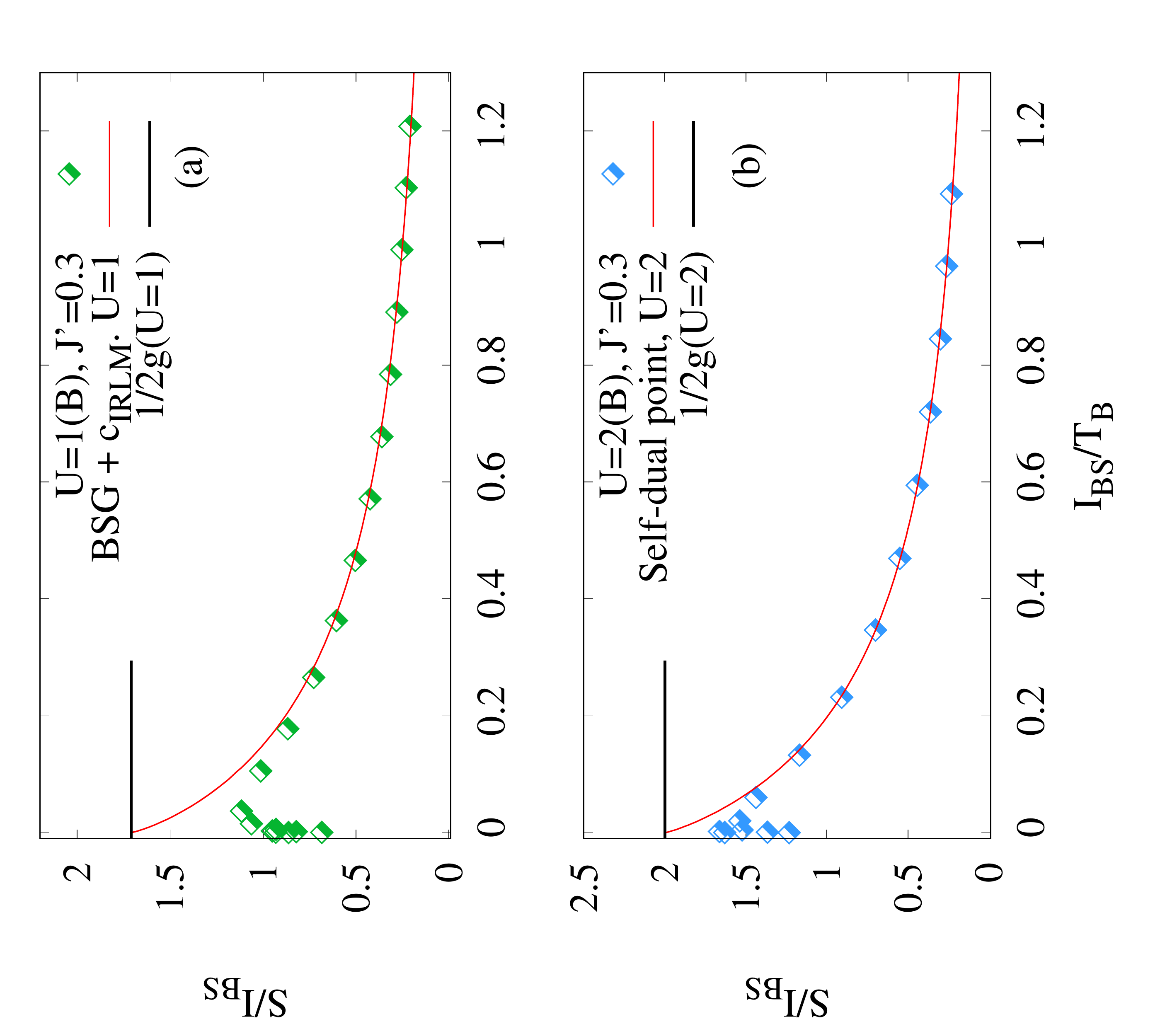}
\caption{$S/I_{\rm BS}$ versus rescaled bias $V/T_B$ for $U=1$(a) and $2$(b).
This ratio is also known as the backscattering Fano factor. Here we used an MPS truncation parameter $\delta=10^{-10}$ and a total simulation time $T=90$ to increase a precision of numerics in the small $V$ regime.}
\label{fig:fano}
\end{figure}

We checked numerically that we get the correct charge ${e\over 2}$ at large voltage, independently of the interaction.
This was done by computing the ratio $S/I$, and verifying that it approaches $e/2$ at large bias.
It is, however, more difficult to extract the charge at small voltage in general. 
Since both $S$ and $I_{\rm BS}$ become very small at low bias (almost perfect transmission), it is difficult to achieve a good numerical precision
for these two quantities and for their ratio - the so-called backscattering Fano factor.
The Fano factor is plotted in Fig.~\ref{fig:fano} for $U=1$ (upper panel) and $U=2$ (lower panel).
Since errors (due to finite-time simulations) are the largest at low bias, one may discard the 2 lowest-bias data points.
In that case, the Fano factor extrapolates to some value close to ${e\over 2g}=2e$ at the exactly solvable point $g={1\over 4}$ ($U=2$),
in agreement with the arguments given in Sec.~\ref{ssec:IR}.
It should be noted that a very similar result has been obtained in Ref.~\cite{branschadel_shot_2010}.
The data for the other values of the interaction are, unfortunately, harder to analyze, but they are  
compatible with a charge  $e$ for all other values of the interaction (as expected from the field theoretic discussion). At $U=1$ for instance (top panel of Fig.~\ref{fig:fano}),
the low-bias limit of the Fano factor is indeed close to 1 (that is $e$) if we again allow ourselves to discard the two points at low $I_{\rm BS}$, where we know -- by comparison with the $U=2$ case -- that the error should be the largest.

\smallskip
\section{Summary and conclusions}

In conclusion, the formulas for the BSG model provide excellent approximations to the steady transport properties of the IRLM, especially for moderate values of the interaction ($U\lesssim3$) strength. While it is tempting to speculate that maybe the two models are in the same universality class out-of-equilibrium, a careful study of subtle aspects such as the tunneling charge  (and consequently the  backscattered current) show that this cannot be the case.
In particular, our analysis of the operators involved in the charge transport shows that the effective charge is $e$ in the IR regime of the IRLM (for all $U\ne U_{\rm sd}$), at variance with the behavior of the BSG model. 

While one can argue that the properties ``must be close" since, after all, the two models coincide exactly for two values of $U$ ($U=0$ and $U_{sd}$), {\sl the remarkable agreement of their I-V and shot noise curves in the crossover regime remains very surprising}.  It suggests in particular that, if the IRLM is integrable out-of-equilibrium, its solution should share many common features with the one of the BSG model. A possible direction of attack to understand better what is going on is to develop a formalism where the IRLM would appear as a perturbation of a BSG model, and study the effects of this perturbation on the scattering of quasi-particles, along the lines e.g. of  \cite{ControzziMussardo}. Another direction is to revisit the open Bethe-ansatz formalism  of \cite{andrei} by focussing on possible similarities/differences with the BSG model. We hope to get back to this question soon.

\section*{Acknowledgements}
We thanks S. Carr and P. Schmitteckert for their useful comments on the manuscript.
H. Saleur also thanks E. Boulat, S. Carr and P. Schmitteckert for many discussions and collaborations on related topics.
This work by supported in part by the ERC Advanced Grant NuQFT.
K. Bidzhiev would like to thank Azamat Elkan for numerous fruitful discussions.

\appendix

\section{Second charge cumulant and zero-frequency current noise}
\label{sec:C2_S}

To establish the (classic) relation between the rate of the second charge cumulant [Eq.~(\ref{eq:S})] and
the current noise, one starts by writing the charge in right lead as 
an integral of the current
\begin{equation}
 \hat Q(t)=\hat Q_0+\int _{0}^t \hat I(\tau)  d\tau. 
\end{equation}
If we denote by ${\rm var}[\hat X]$ the variance $\langle \hat X^2\rangle-\langle \hat X  \rangle^2$ of an operator $\hat X$,
we have
\begin{eqnarray}
C_2(t)&=&{\rm var}[\hat Q(t)]={\rm var}\left[\hat Q_0+\int _{0}^t \hat I(\tau)  d\tau\right] \nonumber \\
&=&{\rm var}\left[\hat Q(0)+\int _{0}^t \Delta \hat I(\tau) d\tau\right]. \label{eq:var}
\end{eqnarray}
In the last equality we have used $\Delta \hat I(t)=\hat I(t)-\langle \hat I(t) \rangle$.
Expanding Eq.~(\ref{eq:var}) we get
\begin{eqnarray}
C_2(t)&=&{\rm var}[\hat Q(0)]+\int _{0}^t\int _{0}^t \langle \Delta \hat I(\tau) \Delta \hat I(\tau') \rangle d\tau d\tau' 
\nonumber \\
&&+ \int _{0}^t \langle \Delta \hat I(\tau) \hat Q(0) \rangle
+ \int _{0}^t \langle \hat Q(0)  \Delta \hat I(\tau) \rangle.
\end{eqnarray}
We then make the assumption that correlator $\langle \hat Q(0) \hat \Delta I(\tau) \rangle$
decays sufficiently quickly with $\tau$, such that the last line in the equation above is small compared
to $t$ when $t\to\infty$. We further assume that 
$\langle \Delta \hat I(\tau) \Delta \hat I(\tau')\rangle$	
decays sufficiently quickly with the time difference $|\tau-\tau'|$.
In the limit $t\to \infty$, the double integral will be dominated by $\tau$ and $\tau'$ of the order of $\mathcal{O}(t)$, and
$|\tau-\tau'|\ll t$. At sufficient large times the system
is in a (quasi) steady state and  two-time correlations only depend on the time difference $\tau-\tau'$. It follows
that the double integral can be approximated by
$t \int _{-t}^t \langle  \Delta \hat I(0) \Delta \hat I(\tau) \rangle d\tau$,
or even by $t \int _{-\infty}^\infty \langle  \Delta \hat I(0) \Delta \hat I(\tau) \rangle d\tau$.
We finally get
\begin{eqnarray}
C_2(t)\simeq t \int _{-\infty}^\infty \langle  \Delta \hat I(0) \Delta \hat I(\tau) \rangle d\tau
\end{eqnarray}
and 
\begin{eqnarray}
S\simeq\int _{-\infty}^\infty \langle  \Delta \hat I(0) \Delta \hat I(\tau) \rangle d\tau.
\end{eqnarray}

\begin{figure}[h!]
\includegraphics[height=0.5\textwidth, angle =270]{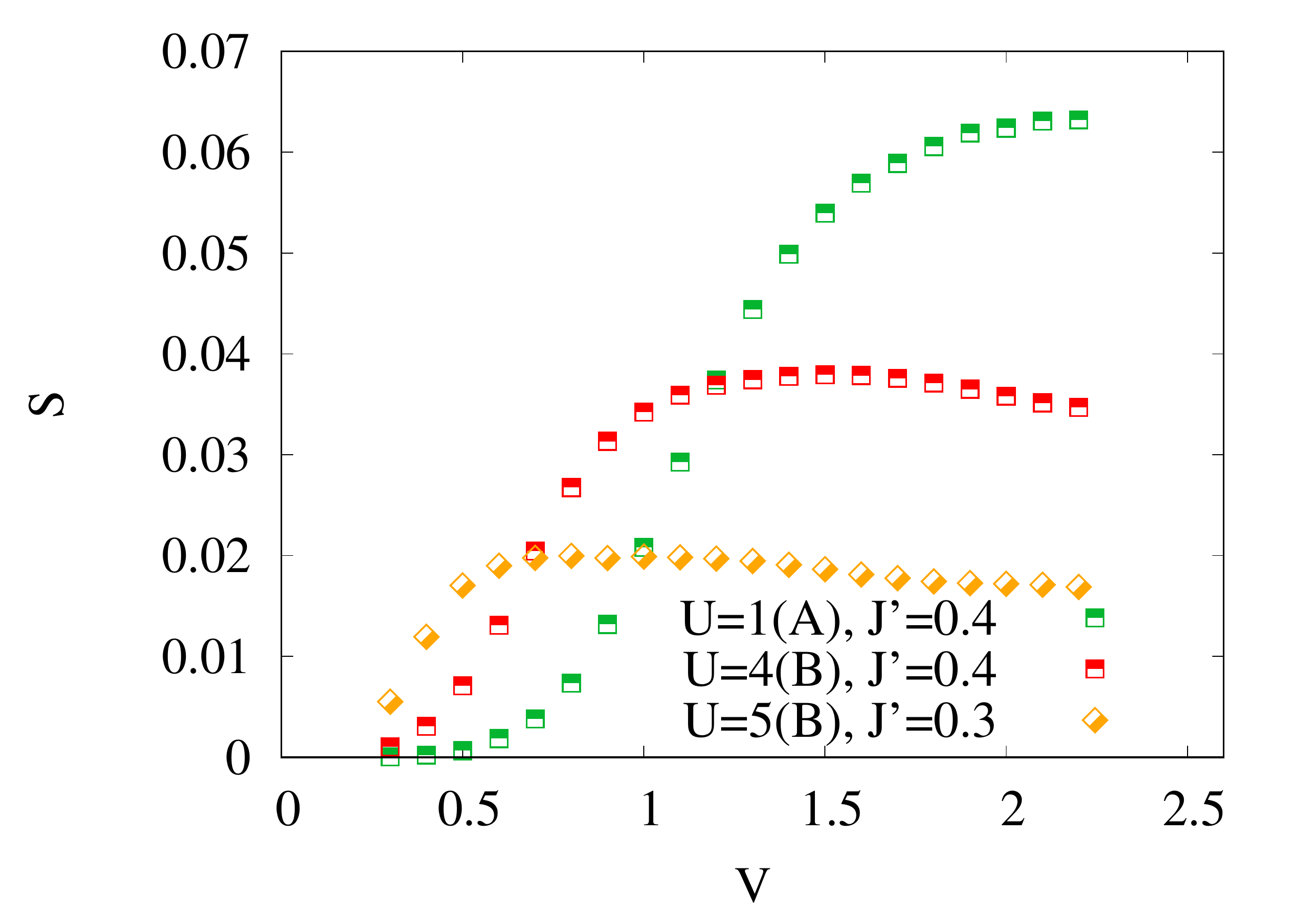}
\caption{Shot noise $S$ as a function of  bare bias $V$ for $U=1(A), 4(B)$ with $J'=0.4$ and $U=5(B)$ with $J'=0.3$.  Most of the calculations are done with $0.6\leq V\leq 1.6$. The low bias $V$ regime is hard to access in this approach since the period of the oscillations $T=4 \pi/V$(see Fig.~\ref{fig:noise-time}) can exceed the simulation time.}
\label{fig:Noisebias}
\end{figure}

\section{Details about the numerical simulations and scaling regime}
\label{sec:numerics}

The simulations are performed using a tDMRG algorithm \cite{white_real-time_2004,daley_time-dependent_2004}
implemented using the C++ iTensor library~\cite{itensor}. We approximate the evolution operator by a matrix-product operator (MPO)~\cite{zaletel_time-evolving_2015} with a 4-th order~\cite{bidzhiev_out-equilibrium_2017} Trotter scheme. The largest time for our numerics is typically $t\simeq40$ with time step $\tau=0.2$, while the system size is $N=257$ sites (128 sites in each lead).

\begin{figure}[H]
\includegraphics[height=0.5\textwidth, angle =270]{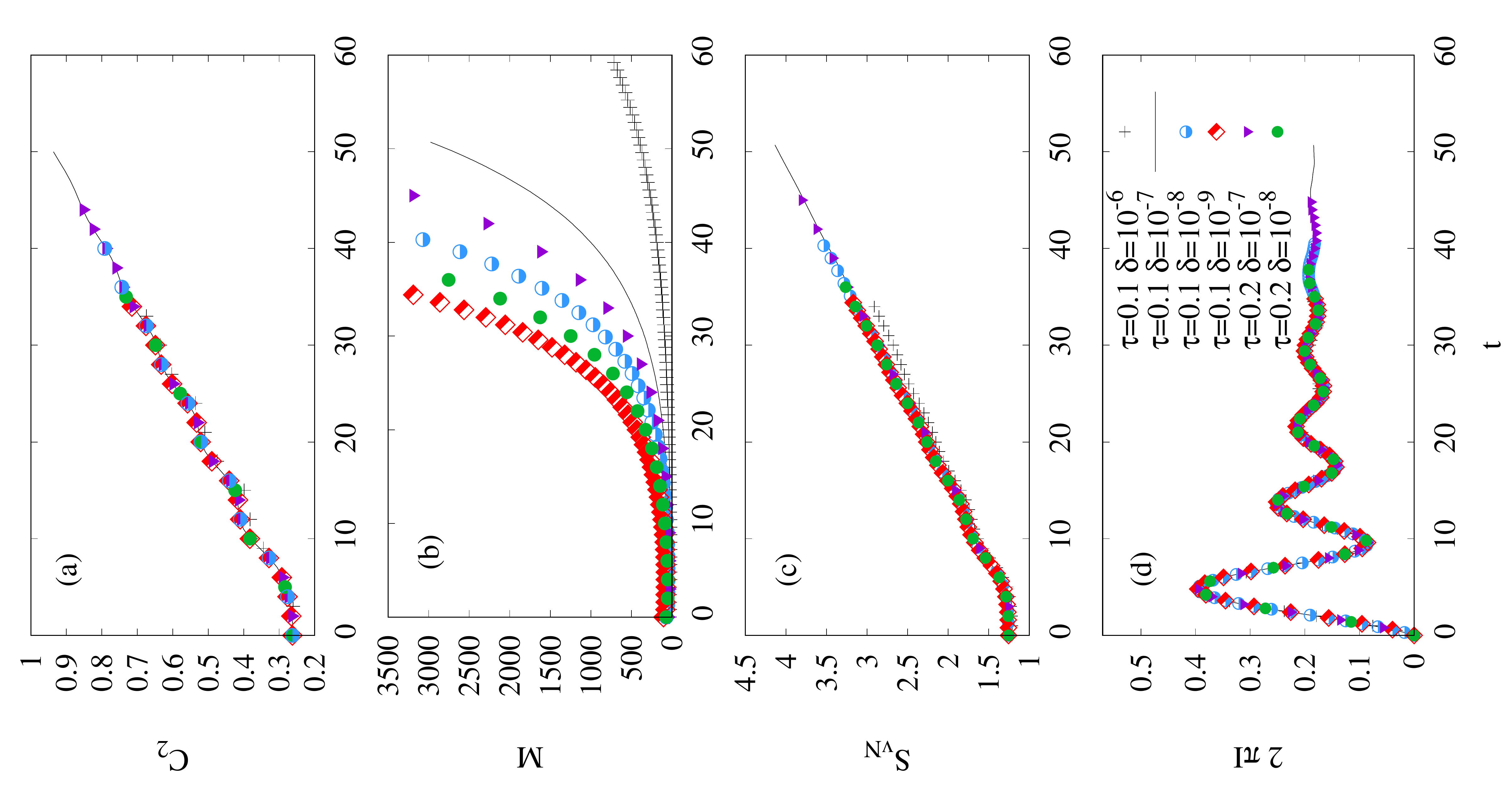}
\caption{Panel (a): second charge cumulant $C_2$ as a function of time and for different values
of the Trotter step $\tau$ and truncation parameter $\delta$ (see bottom panel for the legend). Panel (b): MPS bond dimension $M$ (link between the site $r=-1$ of the left lead and the dot at $r=0$).
In most calculations, the simulation is stopped when $M$ reaches $4000$. Panel (c): von Neumann entanglement entropy $S_{\rm vN}$ between the left and the right leads. Panel (d): Current $2\pi I$ as a function of $t$. Parameters of the model: $N=257, U=6(B), J'=0.3$ and $V=1.6$.
As far as the current or the entanglement entropy are concerned, all the simulations agree relatively well, apart from the less accurate one, with $\delta=10^{-6}$ (crosses).}
\label{fig:tau_delta_dependence}
\end{figure}

The convergence of the data with respect to the maximum discarded weight $\delta$ and Trotter time step $\tau$ is illustrated in
Fig.~\ref{fig:tau_delta_dependence}.
It appears that values between $0.1$ and $0.2$ for $\tau$, and $\delta\sim 10^{-7}$ or below give
good results in the time window we considered. 
As for the bottom panel also shows that an estimate of the steady value of the current can be obtained by fitting $I(t)$
to a constant plus exponentially decaying oscillations (at frequency $f=V/(4\pi)$~\cite{wingreen_time-dependent_1993,branschadel_conductance_2010}).

We are interested in the scaling regime where, in principle, $J'\ll J=1$. However, if $J'$ becomes very small 
the time to reach a (quasi-) steady state becomes very large, which is difficult to handle in the simulations.
In practice we use $J'$ from 0.1 up to 0.5 and $V\lesssim 1.6$ (to be compared with the bandwidth $W=4$).
A large range of $V/T_B$ can then be scanned by varying $V$ and $J'$ in the intervals above.
To check that the model remains sufficiently close to the scaling
regime, one verifies that rescaled quantities, like $I/T_B$, do not significantly depend on $J'$ once they are plotted as a function of the rescaled bias $V/T_B$.

\begin{figure}[H]
\includegraphics[height=0.5\textwidth, angle =270]{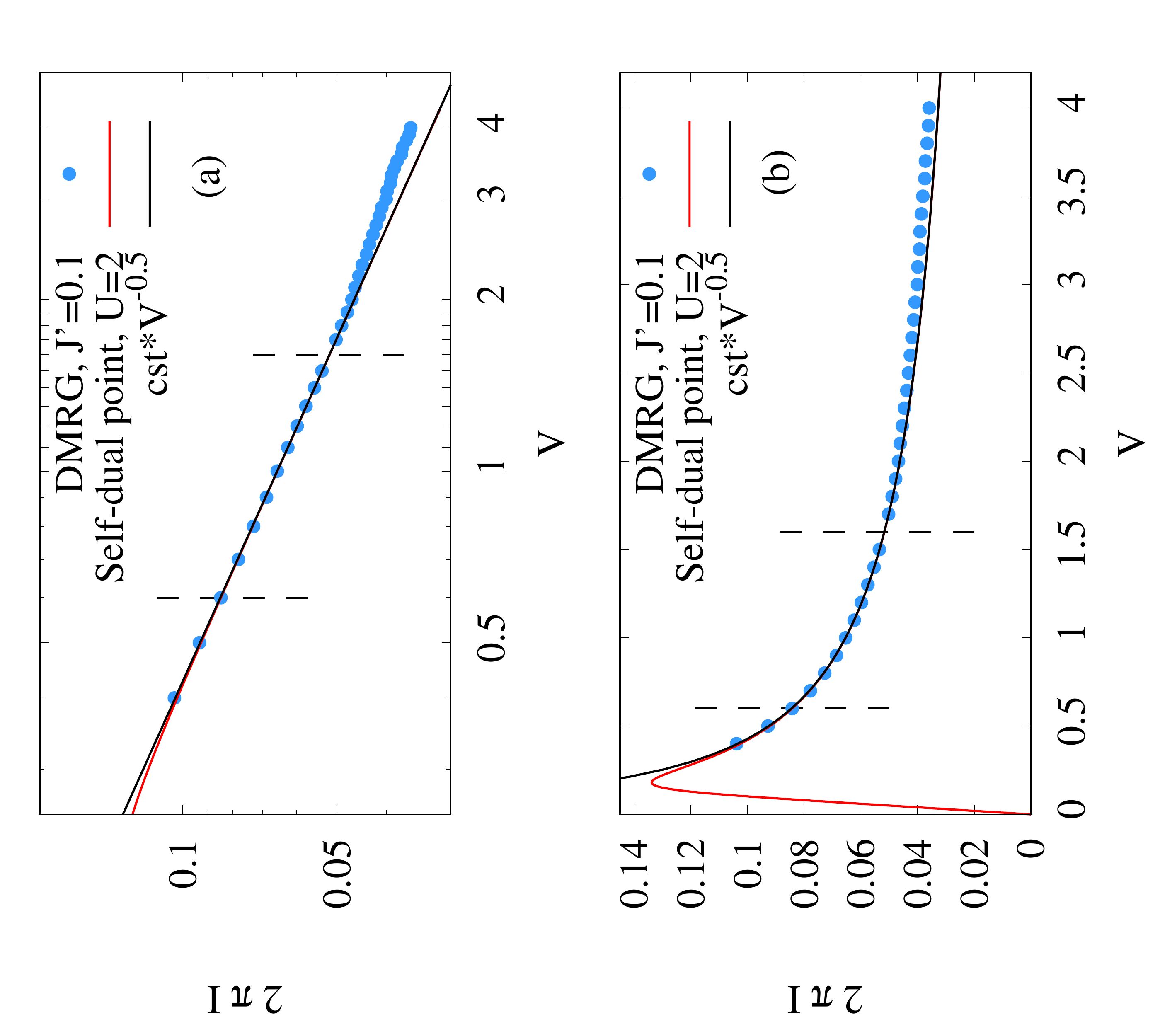}
\caption{Steady current $I$ at $U=2$ plotted as a function of the (bare) bias $V$. The blue dots are tDMRG results
at $J'=0.1$ while the red line is the exact result in the continuum (scaling regime).
Top (a): log scale showing that the current of the lattice model obeys the same power law behavior ($I\sim V^{-\frac{1}{2}}$)
as the continuum limit up to $V\lesssim2$. The vertical dashed lines indicate the bias range used in the present work. Bottom (b): same data in linear scale.
}
\label{fig:V_scaling_regime}
\end{figure}

But to be more precise we investigate below, at $U=2$, 
the values of $V$ and $J'$ beyond which deviations from the scaling regime become visible in the numerics.
The Fig.~\ref{fig:V_scaling_regime} shows the current computed for $V$ up to $4$, and fixed  $J'=0.1$. The tDMRG result
is compared  with the exact result in the scaling regime (red curve). As expected, the agreement is very good at low bias, but it persists up to (almost) $V\sim 2$. Above this
the lattice effects begin to affect the current. For this reason we typically work with $V$ of the order of unity, avoiding too small values which
would cause slow oscillations and difficulties to estimate the asymptotic steady values.

\begin{figure}[H]
\includegraphics[height=0.5\textwidth, angle =270]{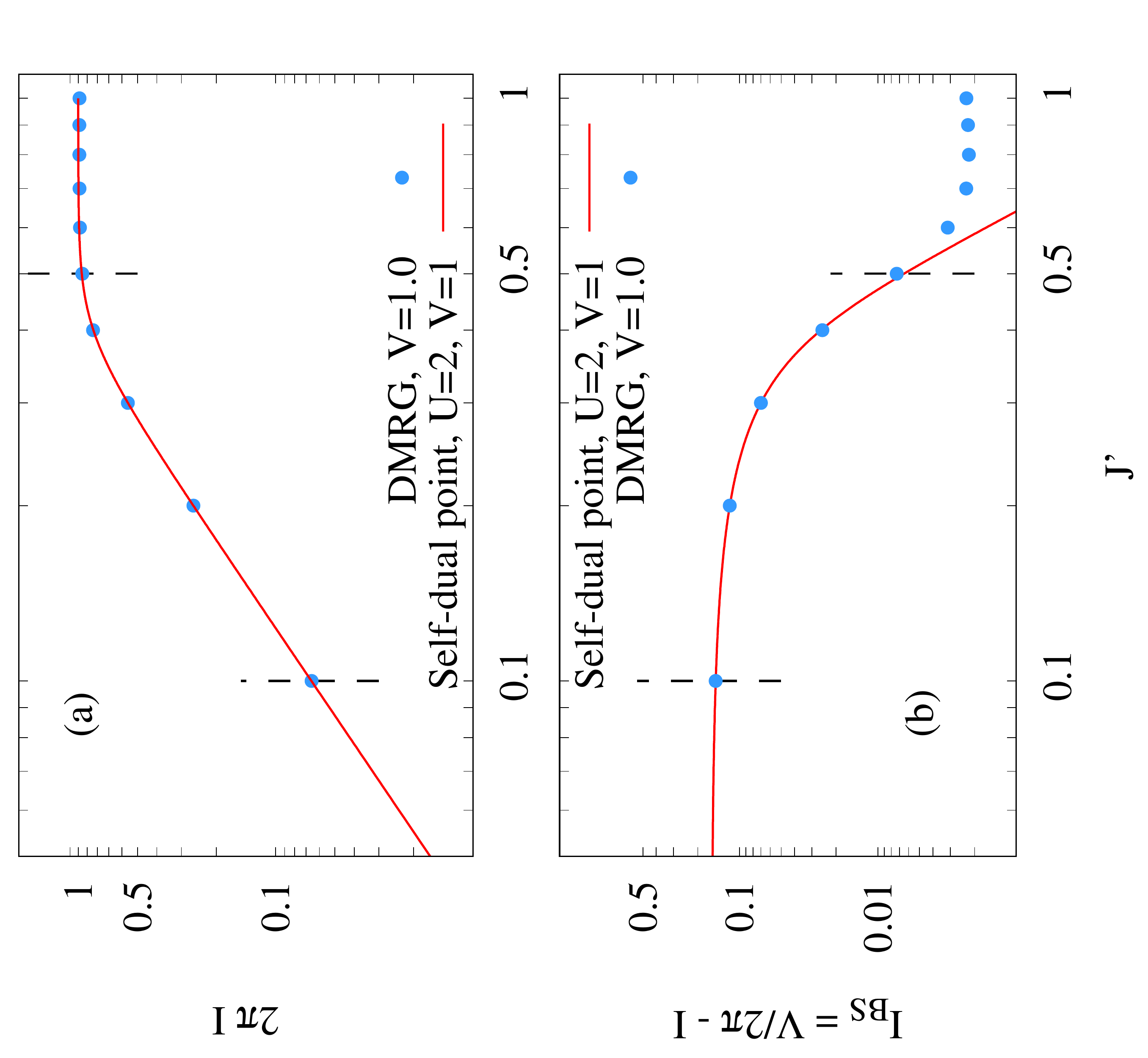}
\caption{Top (a): Steady current $I$ at $U=2$ plotted as a function of the hopping $J'$ and fixed $V=1$ in a log scale.
The vertical dashed lines indicate the range of $J'$ used in the present work.
Bottom (b): backscattered current $I_{\rm BS}=V/2\pi-I$ for the same parameters. 
The comparison with the exact result in the scaling regime (red curve) shows the lattice model is reasonably well described
by the continuum limit up to $J'\simeq 0.5$.}
\label{fig:J_scaling_regime}
\end{figure}

A similar analysis, shown Fig.~\ref{fig:J_scaling_regime}, can be done concerning $J'$. Here again we compare the lattice calculations, at $U=2$,
with the exact continuum result. Both the steady current $I$ and the backscattered current  $I_{\rm BS}=V-2\pi I$
are plotted in Fig. ~\ref{fig:J_scaling_regime}. Looking at the current (upper panel), one may say that we have a good agreement for all the values of $J'$, up to the largest
one (here $J'=1$). But this is misleading: increasing $J'$ at fixed $V$ makes the rescaled bias $V/T_B$ smaller, and we thus go into the IR regime.
In this regime of small rescaled bias and almost perfect transmission the current is essentially given by $I\simeq V/(2\pi)$, even in presence of strong lattice effects. So, to check that the nontrivial features
of the continuum limit of the IRLM model are captured in the simulations, one should look at the deviations from $I\simeq V/(2\pi)$, that is one should analyze the backscattered current.
This is represented in the bottom panel of Fig.~\ref{fig:J_scaling_regime}, and deviations from the scaling regime clearly appear only beyond $J'\simeq 0.5$.
This justifies the fact that in the present study we use $J'$ up to 0.4 or 0.5, while keeping very small the deviations from the scaling regime.


\section{Comparison of the two protocols}
\label{sec:compare_protocols}

As described in Sec.~\ref{ssec:protocols}, two initial states are considered in this study. In the protocol (A) the initial state is constructed as the ground-state of a free fermion Hamiltonian
with homogeneous hopping ($J'_0=J=1$), and a chemical potential bias between the left and the right halves of the chain.
As for  protocol (B), the initial state is constructed as the ground-state of the IRLM
Hamiltonian [Eq.~(\ref{eq:H_IRLM}), with $J'_0=J'<J$ and $U_0=U\ne 0$], to which the bias term [Eq.~(\ref{eq:tanh})] is added.

In the protocol (A) the initial state is built from an Hamiltonian which is spatially homogeneous, apart from the slowly
varying $H_{\rm bias}$.
For this reason, it produces a relatively smooth spatial variation of the fermion density in the vicinity of the dot,
by reducing possible Friedel-like oscillations.
This also produces some smooth variation of the particle density in each lead as a function of the bias, minimizing discretization effects that are present if starting from disconnected (or almost disconnected) finite-size leads~\footnote{These discretization effects are present in the protocol (B) when $J'$ is small.}. 

\begin{figure}[H]
\includegraphics[height=0.48\textwidth, angle=270]{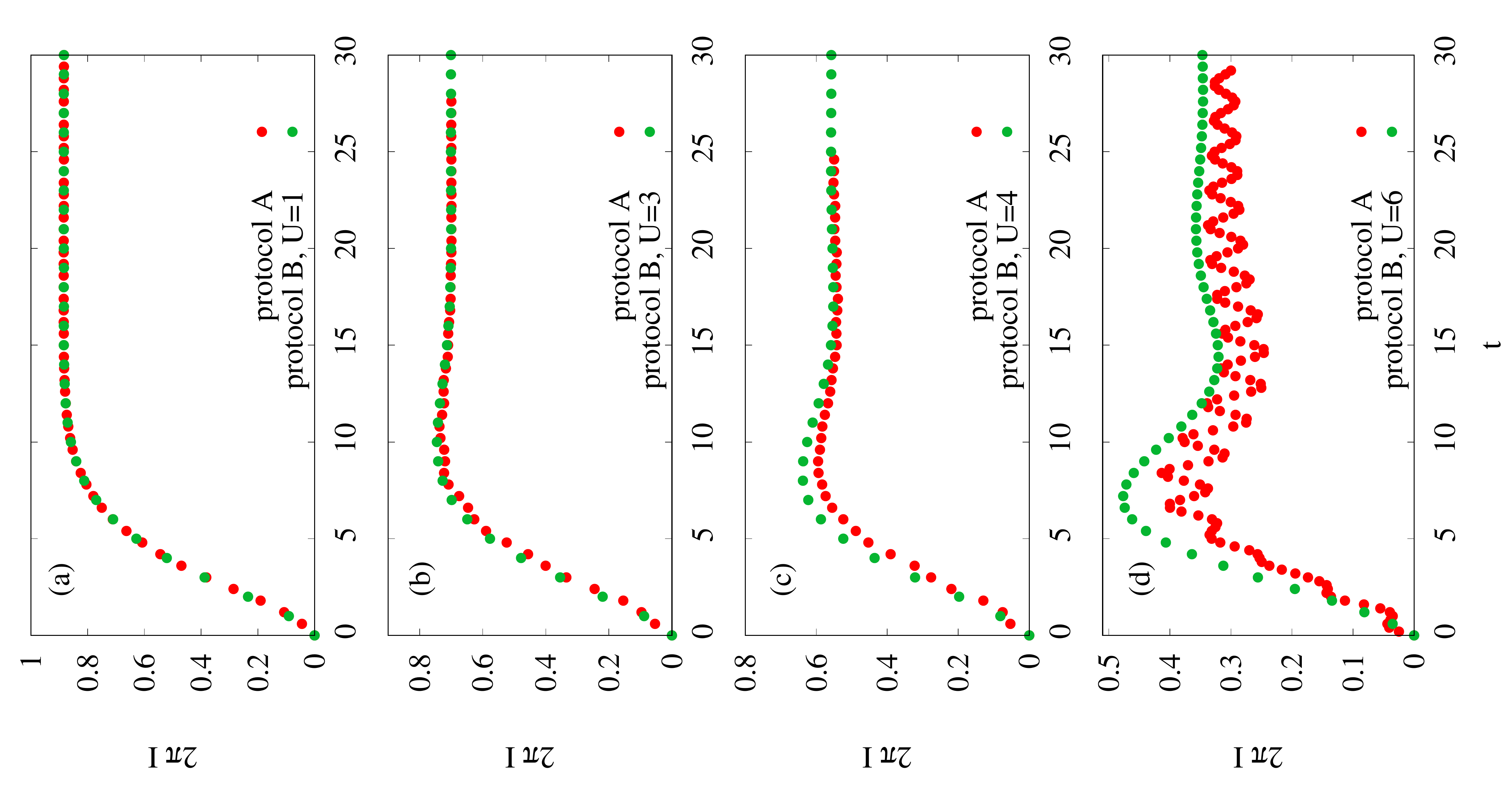}
\caption{Comparison of the current evolution using the protocols (A) and (B). Parameters of the model: $N=257, J'=0.4$ and $V=1.0$.}
\label{fig:QuenCurrent}
\end{figure}

\begin{figure}[H]
\includegraphics[height=0.48\textwidth, angle=270]{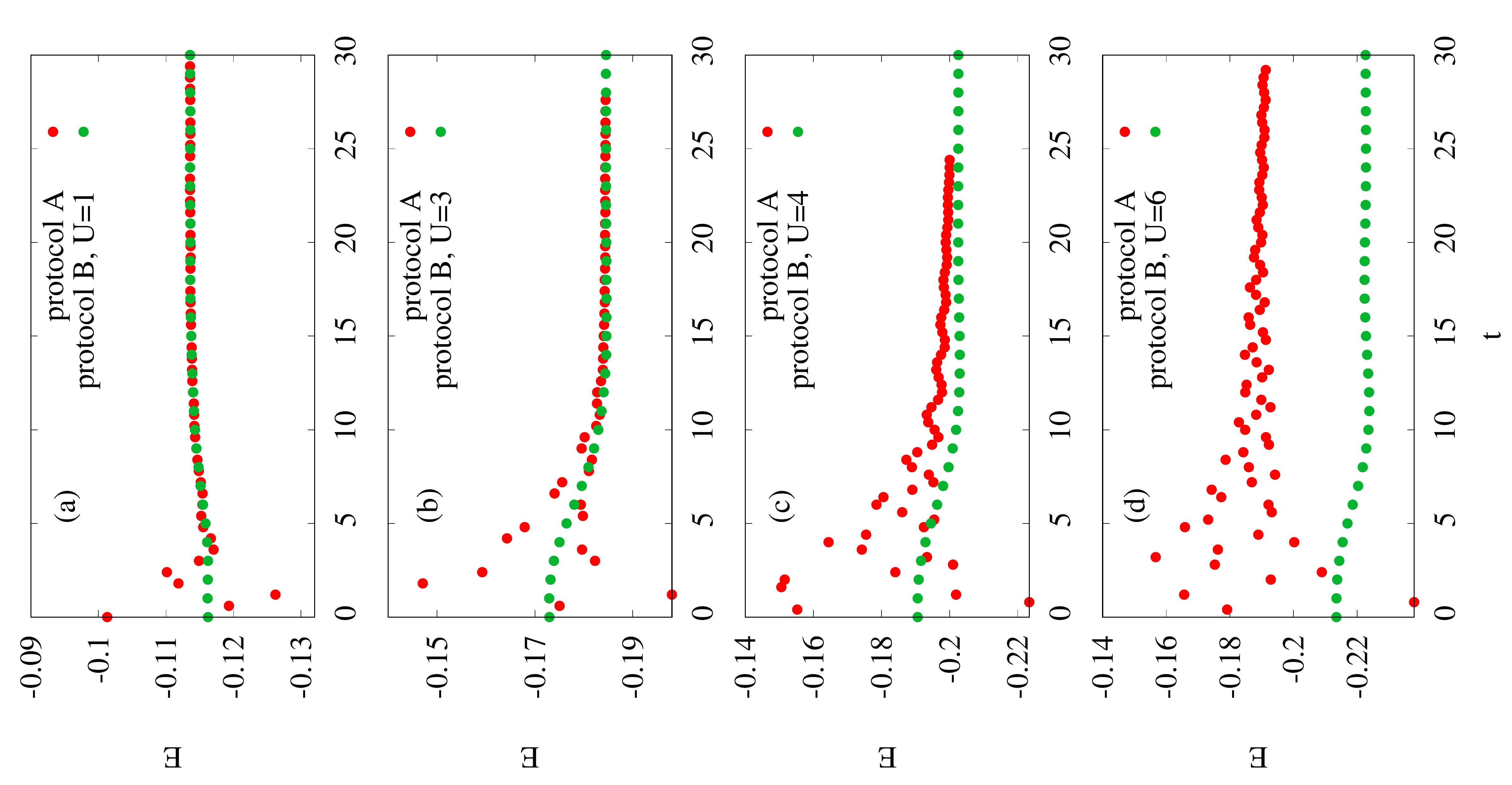}
\caption{Comparison of the average interaction energy on the dot $E=0.5\langle S^z_{-1}S^z_0+S^z_{1}S^z_0\rangle$ evolution using the protocols (A) and (B).
Same parameters as in Fig.~\ref{fig:QuenCurrent}.}
\label{fig:Quenches}
\end{figure}

But since $U$ is switched on at $t=0$ in  (A), the system has some excess energy of order $\mathcal{O}(U)$ that is localized in the vicinity of the dot at $t=0^+$.
This is simply because $\frac{1}{2}\langle S^z_{-1}S^z_0+S^z_{1}S^z_0\rangle$ is lower in the ground state of a model with $U>0$ than in the ground state of a model with $U=0$.
Although this energy is expected to get gradually diluted across the system along the time evolution, we observe that for $U\gtrsim3-4$ it can modify the observables in the vicinity of the dot~\footnote{We thank Peter Schmitterckert for pointing out this effect to us.}.
This is illustrated in Fig.~\ref{fig:Quenches}, where the evolution of $E(t)=\frac{1}{2}\langle S^z_{-1}S^z_0+S^z_{1}S^z_0\rangle$ is displayed as a function of time
in the protocols (A) and (B) and a few values of $U$ from 1 to 6. For $U=1$ or $U=3$, after some transient regime $E(t)$ appears to have essentially the same limit in the two protocols. The value of the steady current  is then also the same (Fig.~\ref{fig:QuenCurrent}). But the situation is different for large values of $U$.
For $U=4$ the interaction energy in the protocol (A) takes a relatively long time to reach that of the protocol (B). In fact, at $t=25$ the protocol (A) still shows some slight excess of interaction energy compared to the case (B). The situation is then quite dramatic for $U=6$, since the up to $t=24$ the interaction energy
in (A) is much higher than that of (B), without any visible tendency to decay to a lower value.
In such a situation it is not surprising that the transport through the dot is significantly different in (A) and (B), as can be seen in the bottom panel of Fig.~\ref{fig:Quenches}, where the protocol (A) cannot be used to estimate the steady current.
Here we expect that much longer times would be needed before the interaction energy that is localized in the vicinity of the dot can ``dissipate'' in the leads in the form of kinetic energy.

\begin{figure}[t]
\includegraphics[height=0.5\textwidth, angle =270]{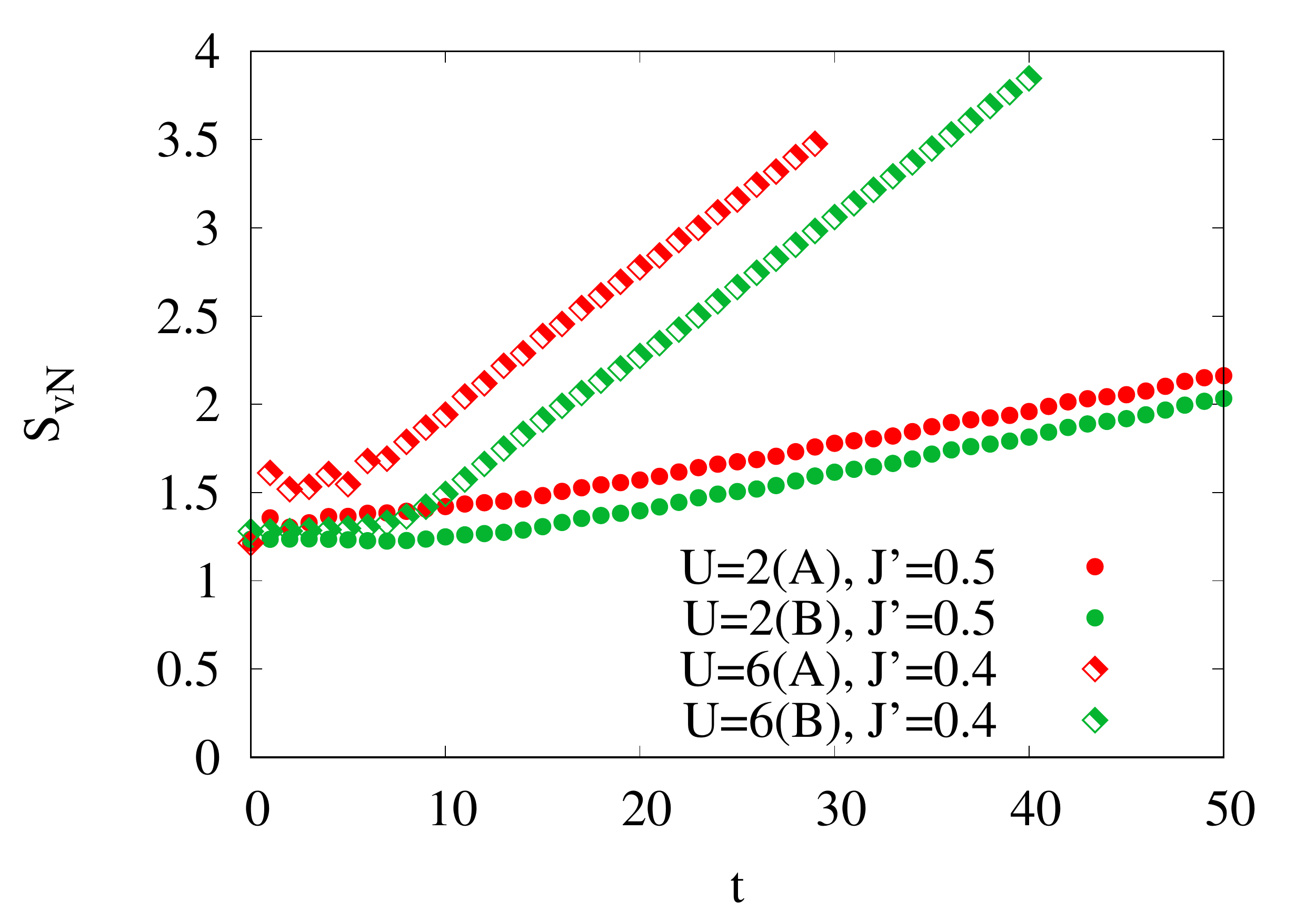}
\caption{Entanglement entropy $S_{\rm vN}$ as a function of time $t$ for $U=2$ and $U=6$ with $V=1.0$. The entanglement entropy is lower with protocol (B) than with (A), especially when $U$ is large. The estimate of the asymptotic entropy rate $\alpha = dS_{\rm vN}/dt$~\cite{bidzhiev_out-equilibrium_2017}  is relatively similar for the two protocols: $\alpha_A(U=6)=0.08$, $\alpha_B(U=6)=0.0788$, $\alpha_A(U=2)=0.0196$, $\alpha_B(U=2)=0.021$.}
\label{fig:entropy-time}
\end{figure}

Finally, we note that the protocol (B) also leads to a smaller entanglement entropy (see  Fig.~\ref{fig:entropy-time}), and for a given maximum bond dimension the simulations can be pushed to longer times.
For these reasons, at large $U$, the protocol (B) where the $U$ term is already  present when constructing $\ket{\psi(t=0)}$ should be preferred.
On the other hand, for $U=0$ and $U=2$, where exact results are available, the protocol (A) appears to give some slightly better results (see  Fig.~\ref{fig:noise-time}).

\begin{figure}[H]
\includegraphics[height=0.48\textwidth, angle=270]{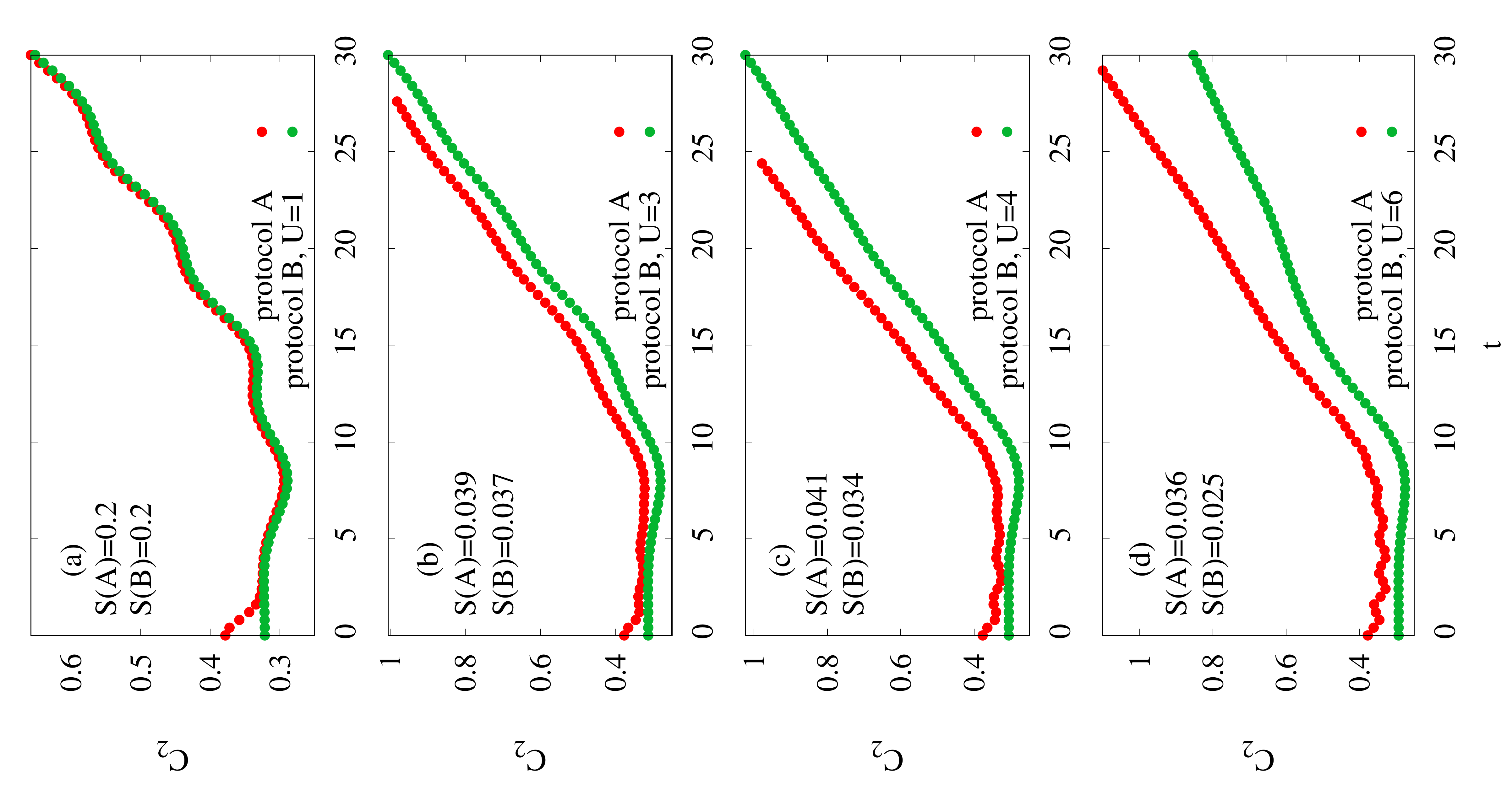}
\caption{Charge cumulant $C_2$ using the protocols (A) and (B). In each case, the estimate of the shot noise $S$ is given in the legend.
For $U\gtrsim3$ the two protocols give markedly different results and in such cases the protocol (B) the most accurate one.
Same parameters as in Fig.~\ref{fig:QuenCurrent}.}
\label{fig:C2_A_vs_B}
\end{figure}

\bibliography{irlm2.bib}{}
\end{document}